\documentclass[format=acmsmall, review=false, screen=true]{acmart}
\usepackage{booktabs} 
\usepackage[ruled]{algorithm2e} 

\SetAlFnt{\small}
\SetAlCapFnt{\small}
\SetAlCapNameFnt{\small}
\SetAlCapHSkip{0pt}
\IncMargin{-\parindent}
\acmJournal{TCPS}
\acmVolume{4}
\acmNumber{2}
\acmArticle{16}
\acmYear{2019}
\acmMonth{9}
\acmArticleSeq{0}
\setcopyright{acmcopyright}
\usepackage[english]{babel}
\usepackage{longtable,psfrag,dsfont,supertabular,bm,units}
\usepackage{amssymb,amsmath}
\usepackage{graphicx} 
\usepackage{algorithmic}
\usepackage{multirow}
\usepackage{multicol}
\usepackage{comment}
\usepackage{subfigure}
\usepackage{fullpage}
\usepackage{url}

\usepackage{xcolor, colortbl}
\usepackage{wrapfig}

\usepackage{listings}
\usepackage{color}
\definecolor{light-gray}{gray}{0.90}
\begin{document}

\title{A System-level Behavioral Detection Framework for Compromised CPS Devices: Smart-Grid Case}

\author{Leonardo Babun}
\orcid{0000-0002-7082-8423}
\affiliation{
  \institution{Florida International University}
  \streetaddress{10555 West Flagler St.}
  \city{Miami}
  \state{FL}
  \postcode{33174}
  \country{USA}}
  
\author{Hidayet Aksu}
\orcid{}
\affiliation{
  \institution{Florida International University}
  \streetaddress{10555 West Flagler St.}
  \city{Miami}
  \state{FL}
  \postcode{33174}
  \country{USA}}
  
\author{A. Selcuk Uluagac}
\orcid{0000-0002-7082-8423}
\affiliation{
  \institution{Florida International University}
  \streetaddress{10555 West Flagler St.}
  \city{Miami}
  \state{FL}
  \postcode{33174}
  \country{USA}}

\begin{abstract}
Cyber-Physical Systems (CPS) play a significant role in our critical infrastructure networks from power-distribution to utility networks. The emerging smart-grid concept is a compelling critical CPS infrastructure that relies on two-way communications between smart devices to increase efficiency, enhance reliability, and reduce costs. However, \textit{compromised devices} in the smart grid poses several security challenges. Consequences of propagating fake data or stealing sensitive smart grid information via compromised devices are costly. Hence, early behavioral detection of compromised devices is critical for protecting the smart grid's components and data. To address these concerns, in this paper, we introduce a novel and configurable system-level framework to identify compromised smart grid devices. The framework combines system and function call tracing techniques with signal processing and statistical analysis to detect compromised devices based on their behavioral characteristics. We measure the efficacy of our framework with a realistic smart grid substation testbed that includes both resource-limited and resource-rich devices. In total, using our framework, we analyze six different types of compromised device scenarios with different resources and attack payloads. To the best of our knowledge, the proposed framework is the first in detecting compromised CPS smart grid devices with system and function-level call tracing techniques. The experimental results reveal an excellent rate for the detection of compromised devices. Specifically, performance metrics include accuracy values between 95\% and 99\% for the different attack scenarios. Finally, the performance analysis demonstrates that the use of the proposed framework has minimal overhead on the smart grid devices' computing resources.
\end{abstract}

\terms{CPS security, Smart Grid security, compromised devices, IEC61850, system calls, function calls.}

\thanks{This material is based upon work partially supported by the U.S. Department of Energy under Award Number DE-OE0000779 and the U.S. National Science Foundation under Award Number NSF-1663051.

Author's address: L. Babun, H. Aksu, and A. S. Uluagac, Electrical and Computer Engineering Department, College of Engineering and Computing, Florida International University; (Current address) 10555 West Flagler St. Miami Fl, 33174.

{\color{red} Version accepted for publication in the ACM Transactions of Cyber-Physical Systems (TCPS), but has not been fully edited. Content may change prior to final publication. FOR EDUCATIONAL PURPOSES ONLY}
}

\maketitle

\section{Introduction}
Critical infrastructure networks such as utility, production, and distribution systems are pillars of any nation and economy. They depend on intelligent and advanced Cyber-Physical Systems (CPS) to guarantee the efficient and reliable delivery of the data generated within these networks. These vital delivery systems have recently been going through a massive effort to modernize their CPS infrastructure. 

In the specific case of the power grid, a substantial effort has already been made to modernize the traditional decade-old grid to the next generation of technology, i.e., \textit{smart grid}. The core concept of the smart grid relies on the integration of the underlying electrical distribution with two-way communications capabilities between the smart CPS devices in the grid. The uses of CPS devices in the grid allows new functionalities and state-of-the-art computing systems for the smart grid infrastructure over the traditional power grid \cite{SEC_grid_comm}. Nonetheless, new security concerns stem from the use of CPS devices by the modern power grid.  

\textit{Indeed, with all its dependency upon device operations and communications, the smart grid is highly vulnerable to any security risk stemming from devices}. Notably, the use of compromised devices can wreak havoc on the smart grid's critical functionalities \cite{SEC_Project, stuxnet2} and can cause catastrophic consequences to the integrity of the smart grid data and operations. Recent examples like the Stuxnet and Sandworm worm attacks \cite{ENISA, attackexamples, reuters} have proven that compromised devices represent a serious threat to the smart grid. Specifically, in the case of Stuxnet, the worm first targeted computers controlling Programmable Logic Controllers (PLCs)), to then change the configuration of the PLCs and cause the uranium centrifuges to behave erratically \cite{stuxnetstory}. The same way, in the case of Sandworm, the attack first targeted computing systems using the BlackEnergy Trojan \cite{black} to gain control over Remote Terminal Units (RTUs) and substation breakers to cause power blackouts \cite{russianhacking}. Due to these real attacks, understanding the behavior of the smart devices, particularly the compromised ones, has become more critical than ever. Several government agencies focus their efforts to protect the critical infrastructure using behavioral-based approaches \cite{russian}.

In this work, we propose a configurable system-level framework to detect compromised devices performing unauthorized operations inside the smart grid \cite{patent, convolution}. Specifically, the proposed framework utilizes system and function call tracing techniques, signal processing, and statistical analysis to detect compromised devices based on their unexpected behavior. In order to test our framework, we designed a realistic representative smart grid substation testbed in which generic CPS devices performed essential operations conforming to the  International Electrotechnical Commission 61850 (IEC61850\footnote{IEC61850 is a protocol suite that defines the communication standards for electrical substation automation systems~\cite{iec61850_1}.})~\cite{iec61850_2, iec61850_3, iec61850_4, iec61850_5} protocol suite. The proposed testbed includes both resource-limited (e.g., RTUs, PLCs, and resource-rich (e.g., Phasor Measurement Units (PMUs), Intelligent Electronic Devices (IEDs)) CPS devices. In the testbed, the devices use open-source $libiec61850$ libraries~\cite{libiec61850} to exchange smart grid time-critical messages using the GOOSE format \cite{iec61850_1}. 

In addition, the adversary model complies with the security requirements specified by the standardization organizations~\cite{NIST_security_panel} for the smart grid. In total, we consider six different types of compromised devices defined by different combinations of device computing resources and attack payloads. 

Finally, we evaluate the performance of our framework by detecting and analyzing behavioral differences between compromised and ground truth devices using three different detection methods. Experimental results demonstrate that the proposed framework achieves an excellent detection rate. Performance metrics reveal accuracy values between 95\% and 99\% for the different types of devices and detection methods analyzed. Additionally, detailed performance analysis shows minimum overhead on the use of the smart grid devices' computing resources (i.e., CPU and memory). On average, memory and CPU utilization does not increase more than 0.03\% and 1.9\%, respectively.

\noindent\textbf{Contributions:} The contributions of this work are as follows:
\begin{enumerate}
    \item {We designed a configurable system-level framework that combines system and function call tracing techniques with signal processing and statistical analysis to detect compromised smart grid devices. To the best of our knowledge, this is the first work that utilizes these techniques in detecting compromised devices in the smart grid.}
    \item {To test the efficacy of our framework, we designed a realistic smart grid substation testbed that included both resource-limited and resource-rich devices. These devices followed a GOOSE publisher-subscriber communication model using open-source $libiec61850$ libraries. {\color{black}The proposed testbed represents a valuable configurable benchmark for this, and other research works on CPS security via behavioral analysis.}}
    \item {In the adversary model, we considered six different types of compromised devices with different computational resources and attack payloads.}
    \item{We evaluated the performance of our framework by detecting behavioral differences between the compromised device and ground truth devices. We obtained accuracy results over 95\% and precision results over 93\% for all the different attacks scenarios and types of devices analyzed. These metrics demonstrated that the proposed framework is highly effective to recognize compromised smart grid devices using behavioral analysis.}
    \item{Finally, our analysis shows that the proposed framework does not represent a significant overhead in terms of computing resources.}
\end{enumerate}

\noindent\textbf{Organization:} The remaining of the paper is organized as follows. Section II presents the background information and some critical implementation assumptions. Then, in section III, we describe the attacker model. Then, in Section IV, we detail the architecture of the proposed framework. In section V, we analyze and discuss experimental results, performance metrics, and benefits of our work. Finally, Section VI presents the related work, and Section VII concludes the paper.


\section{Background and Assumptions}
\label{sec:Background}
{\color{black}In this section, we provide insights into the behavioral analysis of the smart grid devices at the system level.
\subsection{Overview of System-level Smart Grid Substation Architecture}}
{\color{black} The National Institute of Standards and Technologies (NIST) defines the smart grid as a set of seven different interconnected domains \cite{NIST_framework3}. Specifically, two of these domains are responsible for the generation and transmission of electricity, while the other four provide business, operations, and customer support. Finally, at the center of the smart grid architecture, the distribution domain (i.e., smart grid substations) acts as a communication and control hub for the entire infrastructure, which makes it especially attractive to cyberattackers \cite{SEC_grid_comm}.

In Figure \ref{fig:levels}, we present a simplified version of the smart grid distribution domain architecture. Here, three main operation layers can be highlighted \cite{icit, tii, tsg}:
\begin{itemize}
    \item \textit{Process Level}: permits the data acquisition and control at the lowest level of the smart grid substation architecture. The devices at the process level (i.e., merging units) extract state information from sensors, transducers, and actuators and deliver command controls from the upper layers. 
    \item \textit{Bay Level}: permits the two-way communication between the process level and the upper operation layers of the smart grid substations. Here, Industrial Ethernet switches interconnect different control and protection EIDs to allow: (1) protection and control of the data exchanged between bay level and upper and lower layers and (2) protection of the data exchanged between devices located inside the bay level. 
    \item \textit{Station Level}: provides user interfaces and enable applications for engineering and control of the lower layers. Here we can highlight operations from the communication system, the time synchronization system, the substation data collection and control, and servers and workstations. 
\end{itemize}

\begin{figure}[ht]
    \centering{\includegraphics[width=0.75\textwidth]{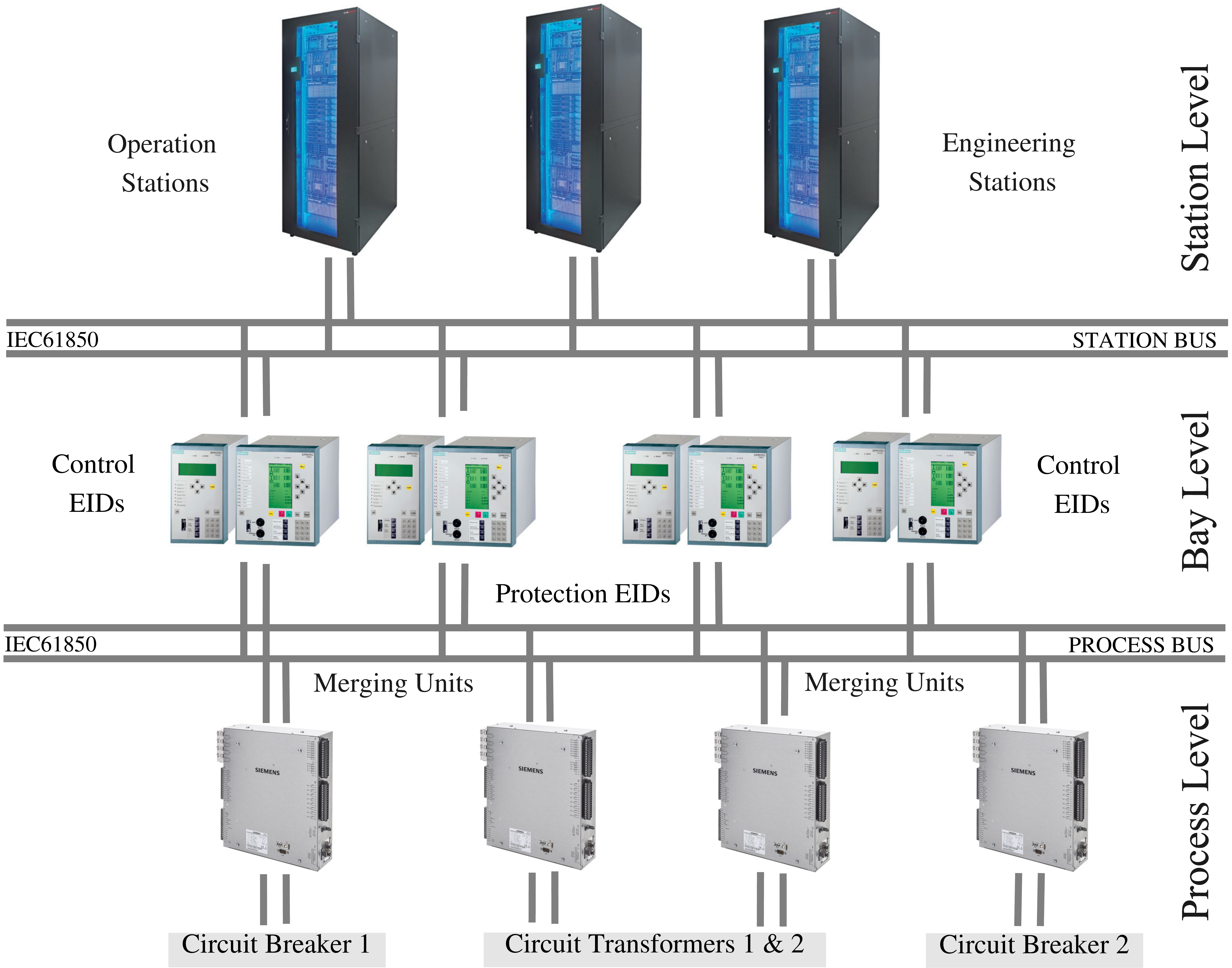}}
    \caption{System-level interaction of smart grid substation devices. The two-way communications under protocol suite IEC61850 can be established both horizontally (between devices from the same level) and vertically (between devices from different levels).}
    \label{fig:levels}
\end{figure}

The IEC61850 protocol suite enables the real-time communications between devices from different substations levels (vertical communications) and devices within the same level (horizontal communications) using Manufacturing Message Specification (MMS), Generic Object Oriented Substation Events (GOOSE), and Sampled Measured Values (SMV) messages \cite{iec61850_1, SEC_enhancing62351}. Specifically, this standard includes many underlying protocol stacks to support and monitor a variety of time-critical services. Indeed, IEC61850 supports real-time operations, abstracts services, and interoperability between devices used in energy automation \cite{iec61850_3, iec61850_2}. 

\subsubsection{Behavioral analysis of substation smart grid devices}
For this work, we focus on the behavioral characteristics of the smart grid substation devices while they communicate and perform either intra- or extra-level operations (i.e., horizontal and/or vertical communications) in the smart grid substation. We define \textit{behavioral characteristics of devices at the system level} as the effect of the device's substations activities on the device's kernel. Let assume that there is a device $O$ performing control operation at the Bay level. These operations can be represented as the set $O_P$ where:

\begin{equation}\label{eq:operations}
    O_P = \{O_{P_0}, O_{P_1}, O_{P_2},...O_{P_N}\},
\end{equation}
then, we define the \textit{system-level behavioral characteristic} of $O$ as the function set $BC$ due to the reflection of $O_P$ at the device's kernel level \cite{syscalls1, syscall2}, that is:

\begin{equation}\label{eq:function}
    BC = f_{kernel}(\{O_P\}),
\end{equation}

In all the cases, we characterize the devices' kernel activity while the devices perform their regular smart grid substation operations. \textit{Indeed, utilizing $BC$ for the compromised device classification allows for a proper generalization of our framework so the proposed solution can also be successfully applied in other CPS domains outside the smart grid.}}

\subsection{Genuine smart grid devices}
We consider a smart grid device as \textit{genuine} when no hardware nor software alteration or tampering has been performed on the device before, during, or after the manufacturing process. To further characterize and identify genuine devices, we define the parameter Index of Likeness (ILI). The ILI computes the similarity between individual operations $O_i$ performed by a single device while executing a specific task $T$ in different time intervals. {\color{black}Similar modeling approaches have been utilized in the literature to characterize CPS \cite{modelling1, modelling2}}. The universe of operations performed by a device to complete a task $T$ at time instant $t=0$ can be defined as:

\begin{equation}
    T(t=0) = \{\cup_{i=0}^{\infty} O_i : \exists O_i \in T\},
\end{equation}

and the value of ILI for different $t$ can then be expressed as:

\begin{equation}\label{eq:ILI}
\rho_{ILI_{t}ILI_{t+i}} = \frac{\sum{O_{t}O_{t+i}-n{\overline{O_{t} O_{t+i}}}}}{ns_{O_{t}}s_{O_{t+i}}},
\end{equation}
where $O_{t}$ represents the set of operations $O_{i}$ performed by the device to complete the task $T$ at the time instant $t$ and $O_{t+i}$ represents the set of operations performed to complete the same task $T$ at the time instant $t+i$. In the same equation, $n$ represents the cardinality of $O$ and $s_{O_t}$ and $s_{O_{t+i}}$ represent the standard deviation of $O$.

Based on our model, a genuine or \textit{ground truth} smart grid device is expected to have a high value of ILI on average. This assumption has been supported in the literature by other research works that characterize Cyber-Physical Systems (CPS) devices (including smart grid devices) as highly deterministic systems \cite{fingerprinting1}. In general, for processes running over time, ILI is expected to take values between 0 and 1: 0 is the result of entirely uncorrelated $O_{i}$s and 1 is the result of remarkably high correlated $O_{i}$s. For a more realistic analysis, our work considers some inherent level of randomness within the device operations. This assumption prevents two $O$s from being completely identical even if one same device performs similar tasks repeatedly over time. 

\subsubsection{Ground-truth Devices}
In the context of this work, ground truth devices constitute particular cases of devices that are known as genuine. We assume full availability to ground-truth devices from every device class present in the smart grid. The proposed framework utilizes these ground-truth devices during its learning process. In the next sections, we define the practical values of ILI that allow for the characterization of ground-truth devices.

\subsection{Compromised smart grid devices} 
The smart grid (and other CPS) devices can be compromised either directly and indirectly. The direct method occurs in cases where the devices are compromised during any of the steps of the supply chain process \cite{supplytransparency, supplychainbook} or via insiders, by \textit{directly} changing the configuration of the devices or their executing apps. Here, the attackers directly target the CPS devices without any other intermediate device. On the other hand, indirect methods are most commonly used and usually require initial access to the computing systems controlling the CPS devices in the network. Once the attacker gains access to those computers, they can change the configuration and behavior of the edge devices \cite{indirectattack}.

We envision that the proposed framework can be utilized to detect compromised devices in both the supply chain and in the field. For that reason, our work considers that genuine devices can be compromised during any stage of the manufacturing and application process. Specifically for our analysis, \textit{we consider a compromised smart grid device as a genuine device with some malicious function installed on it}. The malicious function can be due to compromised hardware or software component~\cite{counterfeithard, compromised2}. Also, the malicious function is expected to change the basic operations of the genuine device. In general, this function can be injected before, during, or after the device's manufacturing process. In Listing \ref{maliciousCode}, we show realistic samples of a compromised device due to code injection. In this specific example, the malicious functions aim to (1) cause degradation on the device's resources and (2) save critical data on a file to be sent later to attackers.

\begin{lstlisting}[caption=Example of malicious code injected to compromised smart grid devices, label=maliciousCode, belowskip=2\baselineskip]
void stress_mem()
{
    srand(time(NULL));
    long size = rand()%2147483647;
    malloc(size);    
}

void save_and_send_later (GooseSubscriber subscriber)
{
    FILE *f = fopen("/root/baduser/data.dat", "a");    
        fprintf(f, "%" PRIu64 "\n", GooseSubscriber_getCriticalValue(subscriber));
        fclose(f);
}
\end{lstlisting}
To further describe the compromised devices, we recall Equation \ref{eq:ILI}. Here, the set of operations $O$ is compromised with a malicious subset $O_{m}$ executed to perform the malicious activity \cite{modelling3}. That is, for compromised devices, the malicious activity impacts the value of ILI by inserting malicious operations $O_{m_i}$ to $O$. Such operations change the device's kernel behavior (Equation \ref{eq:function}) so additional function or system calls are generated (see Listings \ref{sysCallsgood} and \ref{sysCallsbad}). In general, the set $O_{m}$ is expected to follow certain statistical distribution as detailed later in our adversary model. Finally, for compromised devices, Equation \ref{eq:ILI} takes the form:

\begin{equation}\label{eq:ILIm}
\rho_{ILI_{t}ILI_{t+i}} = \frac{\sum{O_{t}O_{m_{t}}O_{t+i}O_{m_{t+i}}-n{\overline{O_{t}O_{m_{t}} O_{t+i}O_{m_{t+i}}}}}}{ns_{O_{t}}s_{O_{m_{t}}}s_{O_{t+i}}s_{O_{m_{t+i}}}},
\end{equation}
where the term $O_{m_{t}}$ represents the malicious operations executed at time $t$ and $O_{m_{t+i}}$ represents the malicious operations executed at time $t+i$.

\noindent\begin{minipage}{.45\textwidth}
\begin{lstlisting}[caption=System calls extracted from a genuine device, label=sysCallsgood, belowskip=2\baselineskip, aboveskip=1\baselineskip]
pthread_detach
malloc
malloc
free
free
signal
malloc
malloc
free
free
.
.
.
.
.
.
\end{lstlisting}
\end{minipage}\hfill
\noindent\begin{minipage}{.45\textwidth}
\begin{lstlisting}[caption=System calls extracted from a compromised device, label=sysCallsbad, belowskip=2\baselineskip, aboveskip=1\baselineskip]
pthread_detach
malloc
malloc
malloc
malloc
open
free
free
signal
malloc
malloc
malloc
malloc
open
free
free
\end{lstlisting}
\end{minipage}

\subsection {Challenges on the behavioral identification of smart grid devices}
Our framework utilizes changes in kernel's behavioral patterns to identify compromised devices. There are three main architectural challenges that our framework needs to overcome:
\begin{enumerate}
    \item Challenge 1: \textit{The device class needs to be considered}. Different types of devices are expected to have different behavior; however, similar devices can also behave differently based on their specific tasks. Such ambiguity can lead to mistakenly identify genuine devices as compromised. For that reason, our framework incorporates (1) device resources (e.g., CPU and memory), (2) type of device, and (3) device task context into the analysis.
    \item Challenge 2: \textit{Device classes are very diverse}. Device class classification would represent an implementation challenge due to the high device diversity present in the smart grid~\cite{NIST_framework3}. Additionally, after the initial classification, the list of devices would need to be checked periodically due to possible changes in network topology or new devices added to the network. 
    \item Challenge 3: \textit{Smart grid devices operations are not fully deterministic}. OS operations possess some degree of randomness that reflects on the device operation list $O$. During the detection process, the framework needs to discriminate between additional operations present in the call lists due to legitimate random processes and real malicious activities.
\end{enumerate}

\subsection{Classes of smart grid devices}
For this work, we group the smart grid devices into different classes. Then, we expect that devices from different classes have different behavior. To correctly group the devices, we consider three main features that address the challenges above: device's computing resource availability, device's type, and device's task context. 

\indent \textit{Resource availability--} we define two different types of devices based on the availability of their computing resources: \textit{resource-rich} and \textit{resource-limited} devices. 
\begin{itemize}
    \item{\textit{Resource-limited devices}: these devices have simple hardware and software architecture. They run with low-performance CPUs and have minimal memory capability. In general, the randomness of the resource-limited devices' kernel behavior highly depends on their software architecture\cite{fingerprinting1}. Also, these devices are built to execute specific tasks inside the smart grid network. Some devices in this group are PLCs and RTUs.} 
    \item{\textit{Resource-rich devices}: these smart grid devices are close in configuration to full-capacity computers. They have a full Operating System (OS), faster multi-core processors, and significantly higher memory than the resource-limited devices. This type of devices executes specialized tasks inside the smart grid network. Some devices in this group are IEDs and PMUs.}
\end{itemize}

Moreover, we group the devices depending on their specific application, brand, and model. For instance, PMUs from the same model and manufacturer can be grouped together while RTUs and PLCs are not considered of the same type. We consider this classification because the devices from different classes have found to behave differently, even if they perform similar tasks.

Finally, the class-classification process of smart grid devices considers the device's task context. For our purposes, the task context involves the type of activity that the devices are performing and their specific logical location inside the smart grid network. That is, we consider that devices of the same type can behave differently if they are handling different types of data from different parts of the network. 

\textit{In general, we consider that the devices perform similar and repetitive tasks over time \cite{fingerprinting1}. Then, our framework takes advantage of this mode of operation to detect compromised devices based on changes in their expected behaviour.}

\subsection{Open-source approach}
Our smart grid testbed utilizes open-source $libiec61850$ libraries~\cite{libiec61850} to exchange smart grid time-critical messages using the GOOSE format \cite{iec61850_1}. The use of open-source software provides some additional design advantages: (1) our solution is more flexible, (2) the framework is more open to customizations which translate on being highly configurable, and finally, (3) our solution can be easily adapted to other open standards which increases interoperability. Therefore, to keep the proposed framework open-source, we implemented our solutions on Linux-based systems. This approach is considered realistic since a very high percentage of smart grid devices still utilize some variant of Unix-based OS \cite{SEC_smart_grid_systems}. We believe that, due to the open-sourced and configurable nature of our testbed, it constitutes an effective benchmark to test the performance of this and other security tools designed to protect the smart grid, that follows the behavioral analysis.

\subsection{Extracting operations from smart grid devices}
We utilize system and function call tracing techniques to extract the set of individual operations $O$ from the devices. These operations are analyzed while the devices perform specific smart grid tasks $T$. We combine function and system call analysis, so the device's activity is detailed from both kernel and application-level, which increases the robustness of the framework. For attackers trying to exploit the calls to stealth their activities, the inconsistencies between system and function calls triggered by the same process can also indicate the presence of malicious activities. We take advantage of the open-sourced Unix-based nature of our testbed to effectively utilize library interposition and ptrace as system and function call tracing techniques, respectively. 

\indent \textit{Tracing system calls with library interposition--} We use dynamic library interposition (LI) since this is a general-purpose system call tracing method that can be applied to most C-compiled programs \cite{hookingJournal}. LI takes advantage of the use of a shared object defined inside the runtime library. This object is in charge of fetching the system calls at the kernel level. At runtime, LI hooks this shared object to intercept the calls and take control of the applications' behavior.   

\indent \textit{Tracing function calls with ptrace--} At the user level, we use Process Trace (i.e., ptrace), a popular Unix-based tool to trace function calls. Ptrace uses an external process that acts as a parent for the C compiled program that wants to be traced. Once the external process attaches to its child, the parent application has full control of every time the traced application makes a function call. 

Finally, for cases where the smart grid devices do not use Unix-based OS (e.g., Real-Time Operating System (RTOS)), similar approaches are utilized to trace the system and function calls. Similar hooking techniques are possible to use because these other systems behave in similar ways as Linux since they are also POSIX-compliant OS. In general, the tracing technique utilized for hooking into the system and function calls is a configurable feature that depends on every specific application \cite{hookingJournal}. 


\section{Adversary Model} 
\label{sec:AdversaryModel2}
Our adversary model considers, conforming to the NIST guidelines, three possible threats in the smart grid that are directly related to the use of compromised devices~\cite{SEC_grid_home}:

\begin{enumerate}
    \item {\textit{Threat 1 (Information leakage)}: the compromised device opens additional communication channels to leak valuable smart grid information to the adversary (another untrusted insider or outsider) in real-time.}
    \item{\textit{Threat 2 (Measurement poisoning)}: the compromised device generates fake data that can be used to poison the real status of the smart grid.}
    \item{\textit{Threat 3 (Store-and-send-later)}: the compromised device stores information in hidden files that are recovered later by an attacker.}
\end{enumerate}

Based on these three well-defined threats and considering both resource-limited and resource-rich smart grid devices, we further define six different types of compromised devices as part of the adversary model:

\begin{enumerate}
    \item {\textit{Compromised Device 1 ($CD_1$)}: the resource-limited device creates additional instances of the IEC61850 GOOSE publisher object and starts leaking information through unauthorized communication channels.}
    \item {\textit{Compromised Device 2 ($CD_2$)}: the resource-limited device allocates small and unauthorized amounts of memory to create fake data and poison real measurements.}
    \item {\textit{Compromised Device 3 ($CD_3$)}: the resource-limited device creates unauthorized hidden files to store critical information which is retrieved later by the attacker.}
    \item {\textit{Compromised Device 4 ($CD_4$)}: the resource-rich device creates additional instances of the IEC61850 GOOSE subscriber object and starts leaking information through unauthorized communication channels.}
    \item {\textit{Compromised Device 5 ($CD_5$)}: the resource-rich device allocates small and unauthorized amounts of memory to create fake data and poison real measurements.}
    \item {\textit{Compromised Device 6 ($CD_6$)}: the resource-rich device creates unauthorized hidden files to store critical information which is retrieved later by the attacker.}
\end{enumerate}

A summary of the adversary model, its impact on device resources, and the targeted security services of such attacks in the smart grid infrastructure are given in Table~\ref{tab:devices}.

\begin{table}[ht]
\begin{center}
\small
\caption{Threats to the Smart Grid Devices.}
\label{tab:devices}
\begin{tabular}{| c | c | c | c |}
\hline
\multicolumn{4}{|c|}{Adversary Model} \\ \hline
\hline
\textit{Name} & \begin{tabular}[x]{@{}c@{}}\textit{CPS Device}\\\textit{resource availability}\end{tabular} & \textit{Computing resources impacted} & \textit{Security services  compromised}\\ \hline
               $CD_1$ & Limited & Memory, CPU, communications & Confidentiality \\ \hline
               $CD_2$ & Limited & Memory, CPU &  Integrity          \\ \hline
               $CD_3$ & Limited & Memory, CPU & Authenticity, confidentiality \\\hline
               $CD_4$ & Rich    & Memory, CPU, communications  & Confidentiality\\ \hline
               $CD_5$ & Rich    & Memory, CPU & Integrity  \\ \hline
               $CD_6$ & Rich    & Memory, CPU & Authenticity, confidentiality\\ 
\hline
\end{tabular}
\end{center}
\end{table}
    
We also assume that the compromised devices perpetrate their attacks following a Poisson distribution. Poisson allows for randomly and efficiently spacing the attacks and constitutes a valid model to emulate the randomness of such events~\cite{RossBook}.

The behavior of the compromised device is modeled as follows. Consider t=$[0,T]$, the communication interval between the two smart grid devices. The probability of having an attack from a compromised device $CD_{i} \in \{CD_1, CD_2, CD_3, CD_4, CD_5, CD_6\}$ can be expressed as:
\begin{equation}\label{eq:Poisson}
P_{cd} = \frac{{\lambda}^k e^{-\lambda}}{k!}, \hspace{6pt}  k \in \mathbb{R},
\end{equation}
where $\lambda$ is the average number of attacks in the interval of time $t$ and $k$ is the total number of attacks in the same interval. 


\section{Overview of the Proposed Framework}
\label{sec:Framework}
\begin{figure}[!ht]
    \centering{\includegraphics[width=0.7\textwidth]{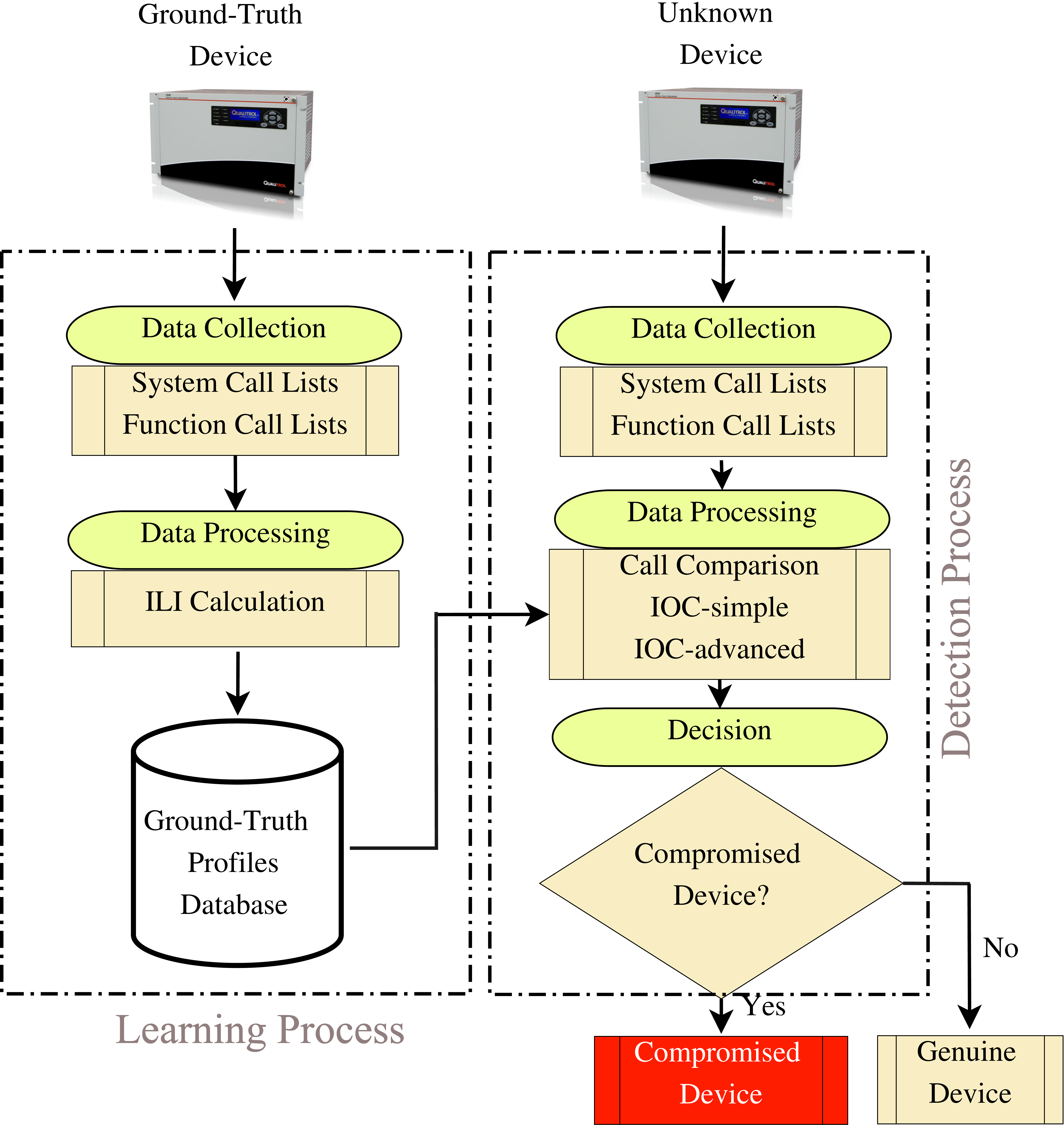}}
    \caption{Configurable framework proposed to monitor and detect compromised smart grid devices. The learning process creates signatures based on ground-truth devices that are utilized later to decide on potentially-compromised devices.}
    \label{fig:framework}
\end{figure}

In this section, we describe the proposed framework to detect compromised devices in the smart grid. Also, we present the details of our detection approaches and decision algorithm. 
Figure \ref{fig:framework} depicts the general architecture of the framework. As discussed before, the main goal of the framework is to decide if an unknown smart grid device is genuine or compromised \cite{ICCPaper}. For this work, the term \textit{unknown} refers to the level of uncertainty regarding the smart grid device being compromised or not. Initially, as part of the \textit{learning process}, a ground-truth device from a specific device class is evaluated to generate its corresponding device-class signature or Ground-Truth Profile (GTP). This signature contains behavioral profiling information from the device and is utilized to decide whether an unknown device from the same class is genuine or compromised. Once the signature is obtained, it is stored in the \textit{Ground-Truth Profiles Database}. In our implementation, we define a separate service to execute the learning process. Such a service separation permits the generation of new signatures every time that new devices join the network. Also, an independent learning process guarantees the replacement of old signatures every time that known devices assume new roles in the smart grid network.

The second part of the framework (also known as \textit{detection process}) starts by extracting a similar profiling signature from the unknown device. Here, we assume that we have enough information to classify the device into some specific device class. Then, three different detection methods are applied to compare and correlate the unknown signature to the corresponding GTP from a similar device class stored in the database. Comparison and correlation results are then used to remove uncertainty and decide if the unknown device has been compromised or not. 

\begin{figure}[!ht]
	\centering{\includegraphics[width=0.60\textwidth]{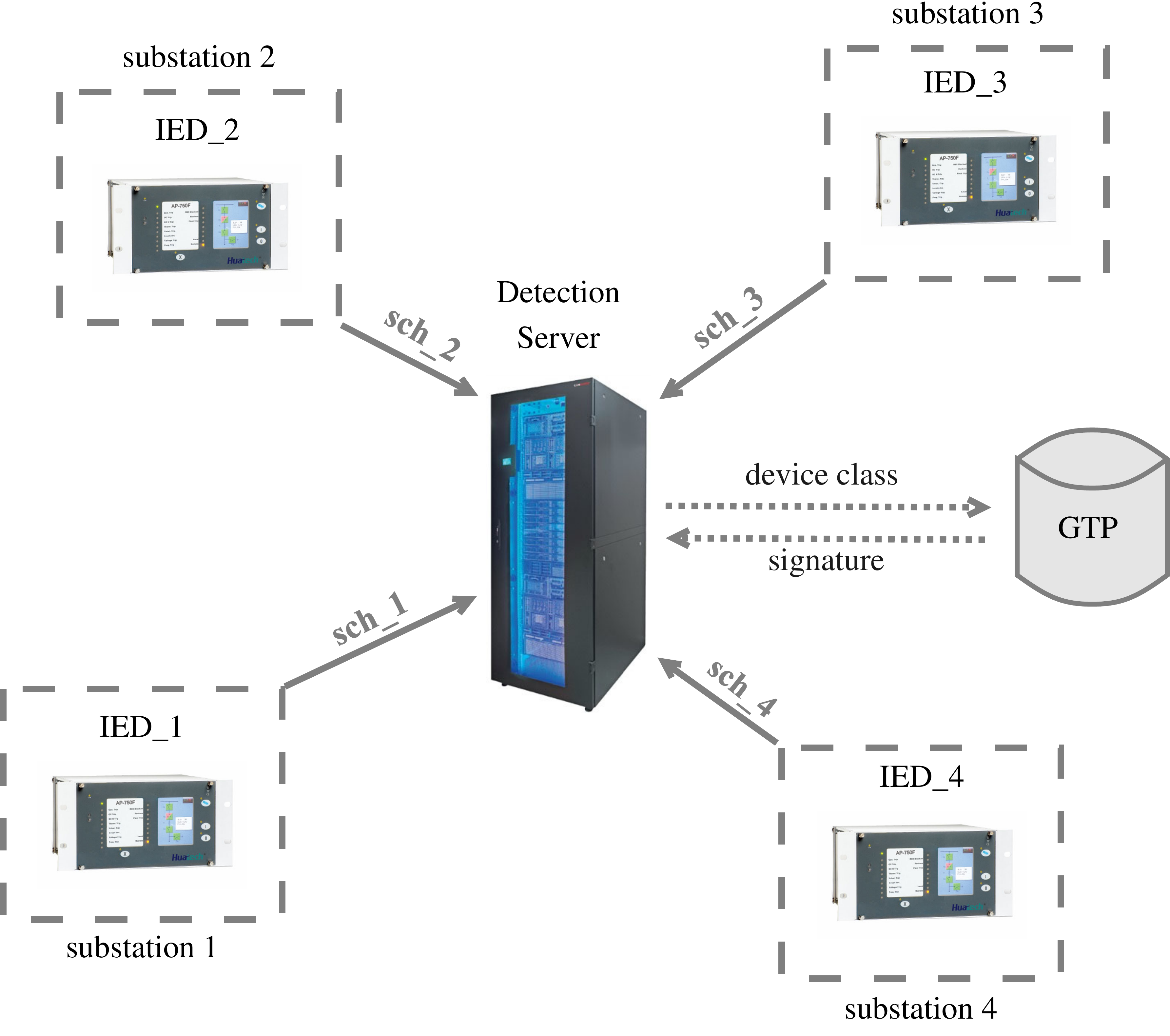}}
	\caption{Example implementation of the proposed framework.}
	\label{fig:implementation}
\end{figure}

We envision the proposed framework as a secured, centralized, and supervised agent virtually located inside the smart grid network. There are several advantages from this implementation model; first, our framework would be compliant with the security challenges of the smart grid \cite{win, wireless, icit}; second, a centralized solution represents a better option to monitor remotely-located devices from different networks; and third, a supervised agent allows for monitoring group of devices without degrading or interrupting critical tasks inside the smart grid. Figure \ref{fig:implementation} depicts a simplified implementation example of our framework. Here, IED devices exchange information between different substation level networks while a detection agent is monitoring them. Inside the devices, a lightweight scheduler (sch) runs parallel processes at the kernel level to hook into the devices' tasks and extract behavioral information. The collected information is sent to the server along with specific device class information using secure TCP-IEC61850 channels via either proxy or VPN-tunnel protected (depending on the smart grid device capability). Then, on the server-side, every scheduled action is processed using either priority or first-in-first-out (FIFO)-based queues. The priority is assigned depending on the device class and may also regulate the frequency of the scheduler's execution. For every detection process, the server executes queries to the GTP database using the device class ID and receiving the corresponding behavioral signature. Finally, the server correlates the scheduler data with the stored signature and decides on the devices as being compromised or not. 

\subsection {Probability of detecting a compromise smart grid device}
In Section \ref{sec:AdversaryModel2}, we presented the probability of having a specific attack during a time interval $t$, considering the device is compromised. In this section, we formally describe the probability of detecting such attacks by using the proposed approach. To generalize, we consider that the statistical relationship between the two discrete random variables $X$ and $Y$ that represent the ground-truth signatures and the timed operation of the unknown smart grid devices follow a bi-variate distribution $B$. From here, we assume that the probability of having a particular specific sequence of calls in the GTP is $P(X)$. The same way, we assume a specific sequence of calls extracted from the unknown device with probability $P(Y)$. 

When an attack occurs, and it is detected, the expected value $E(X)$ and $E(Y)$ of the random variables representing both the GTP and the unknown device call list are $P(X)$ and $P(Y)$, respectively. From here, we can determine the variance $V$ of the attack indicator\footnote{An attack indicator is represented by the value of the random variable that would indicate the presence of an attack.} $\phi_x$ and $\phi_y$ from both the GTP and the unknown call lists as:

\begin{equation}\label{eq:profile1}
Var(\phi_X) = E(\phi_X^2) -  E(\phi_X)^2 = P(X) (1 - P(X)),
\end{equation}

\begin{equation}\label{eq:profile2}
Var(\phi_Y) = E(\phi_Y^2) -  E(\phi_Y)^2 = P(Y) (1 - P(Y)).
\end{equation}

We directly establish the statistical correlation between the random variable X and Y as the co-variance of these attack indicators:

\begin{equation}
\begin{split}
Cov(\phi_X, \phi_Y) = E(\phi_X \phi_Y) - E(\phi_X)E(\phi_Y),\\
                    = E(\phi_{X \cap Y}) - E(\phi_X)E(\phi_Y),\\
                    = P(X \cap Y) - P(X)P(Y).
\end{split}
\end{equation}

Then, we can define the correlation between ground-truth device signatures and the unknown smart grid devices based on the probability of detecting the attacks.

\begin{equation}
\begin{split}
\rho(\phi_X, \phi_Y) = \frac{P(X \cap Y) - P(X)P(Y)}{\sqrt{P(X)(1 - P(X)) P(Y)(1 - P(Y))}},\\
                     = \frac{(P(X | Y) - 1) P(Y)}{\sqrt{P(X)(1 - P(X)) P(Y)(1 - P(Y))}},
\end{split}
\end{equation}
where $P(X | Y)$ represent the conditional probability of detecting an attack on a smart grid device after assuming a ground-truth signature from the same device class has been found. In general, we describe the successfulness of the proposed detection approach to be the jointly bi-variate variable $(X_i, Y_j)$ with probability of occurrence $P(X_i > X_j | Y_i > Y_j)$ for any pair of calls $i$, $j$.   

\subsection{Learning Process}
The primary goal of the learning process is to populate the Ground-Truth Profiles Database that contains all the GTPs from device classes in a specific smart grid network region. The execution of the learning process solves the first two architectural challenges of our framework described in Section \ref{sec:Background}.
The learning process classifies the ground-truth devices into device classes and keeps the GTP database up-to-date. For every different class of devices, the learning process performs two specific tasks: (1) GTP data collection and (2) GTP data processing. \\
\indent \textit{GTP Data Collection--} this stage applies library interposition and ptrace to extract the lists of system and function calls, respectively. These operations are performed while the ground-truth devices execute regular smart grid substation tasks $T$. For every different device class, specific tasks are repeatedly executed over time while the framework hooks every iteration. As a result, for every iteration of $T$, the learning process generates new lists of system and function calls from the ground-truth device. In the end, the data collection process generates a set of system and function call lists. Every list contains detailed information about the specific operations that the devices executed at both the kernel and the user level in every different run of $T$. \\
\indent \textit{GTP Data Processing--} the data processing stage calculates the ILIs for every different ground-truth device class. The concept of ILI introduced in Section \ref{sec:Background} evaluates how much deterministic the performance of a ground-truth device is over time. The more deterministic, the higher the ILI value and the more suitable the ground-truth device is to obtain its GTP. In total, the framework calculates two different values of ILIs, one from the set of system call lists and one from the set of function call lists, respectively.
To successfully calculate the ILIs, the framework assigns a different weight $\delta_{i}$ to every different type of system or function call in the order that they appear. The assignment of $\delta_{i}$ weights constitutes another configurable feature of our framework. This can be done randomly (the weights are considered normally distributed for simple processes where the different system or function calls have the same level of impact on the completion of the task $T$) or by following a specific assignment criterion (adaptive assignment). The adaptive assignment depends on the importance of the specific calls and the type of application that is being evaluated. As a result of the assignment step, the framework generates a random variable $R$ that takes values between $\delta_{min}$ and $\delta_{max}$. This variable describes the behavior of $O$ for every different system or function call list.
Finally, the framework calculates the ILIs using the Equation \ref{eq:ILI}. In the end, the ILI values are compared against a configurable threshold $\sigma$. Initially, the framework selects an initial value for the threshold based on the device class, and then it continues adjusting this value until the average performance reaches the desired target value for that specific class. If both ILI values are above $\sigma$, the GTP is accepted and stored in the database. 

Equation~\ref{eq:profile} represents the general format of the GTP used in our work. The final profile contains information about the device class (DeC),  the entire set of system and function call lists (SCL), and the threshold $\sigma$. At the end of the learning process, the Ground-Truth Profiles Database contains all the possible signatures that characterize the different device classes within a specific smart grid network region.

\begin{equation}\label{eq:profile}
GTP = \{DeC, SCL, \sigma\},
\end{equation}

\subsection{Detection Process}
The main goal of the detection process is to use the profile information stored in the Ground-Truth Profiles Database to determine if the unknown devices are being compromised or not. This process performs three main tasks: 
\begin{enumerate}
   \item  \textit{Data collection}: this step follows almost the same sequence of operations detailed in the learning process. However, this time, the framework obtains the call lists from a single execution of $T$ on the unknown devices from the smart grid networks. No $T$ task is fixed for computing purposes nor is repeated over time.
    \item \textit{Data processing}: constitutes the core of the detection process. In this step, the framework combines three different detection mechanisms to detect compromised devices. The application of every detection approach is decided on-demand, which has a positive impact on reducing the total overhead introduced by our framework.
    \item \textit{Decision}: finally, the decision algorithm processes the results from data collection and processing to decide if the unknown device is genuine or compromised.
\end{enumerate}

In the following, we provide details about the three detection mechanisms.

\subsubsection{Detection mechanisms}
Our framework implements three different detection mechanisms. To utilize computational resources efficiently, the detection mechanisms are applied orderly on-demand. That means, our framework utilizes each detection approach in an ordered fashion, and it always uses the best effort to make a final decision by applying the minimum number of detection steps.\\
\indent \textit{System and Function call list comparison--} The simplest detection approach directly compares the SCL from the GTP to the system and function call lists extracted from the unknown device. The comparison schema considers the type and amount of system and function calls in both GTP and the newly extracted lists. This mechanism is implemented, as shown in Equation \ref{eq:comparison}. Specifically, the comparison approach generates a \textit{call vector} that contains the total number of different calls extracted from the unknown device of class $c$ and normalized against the term $GTP(c; SCL)$. Equation \ref{eq:comparison} details this process:

\begin{equation}\label{eq:comparison}
call\_vector = \{\frac{unkc_0}{GTP_{SCL_0}}, \frac{unkc_1}{GTP_{SCL_1}}, ..., \frac{unkc_n}{GTP_{SCL_n}}\},
\end{equation}
where the term $unkc_0$ refers to the amount of system or function calls of type 0 extracted from the unknown device and the term $GTP_{SCL_0}$ refers to the amount of system or function calls of type 0 extracted from the GTP of the same device class. As inferred from Equation \ref{eq:comparison}, call vector's items of value 0 represent types of calls that are present in the GTP SCL but not in the lists of calls acquired from the unknown device. On the contrary, call vector's items of value $\infty$, represent calls extracted from the unknown device, but that cannot be found in the corresponding GTP. In general, the execution of this first detection approach is very light in terms of computing resources. \\
\indent \textit{Index of Correlation Simple--} A second detection mechanism calculates the statistical correlation between call lists from the unknown device and the GTP. In this case, in addition to the type and amount of calls, the framework considers the order in which these calls are being triggered. The result from calculating such statistical correlation is known as \textit{Index of Correlation simple} (IOC-simple). IOC-simple is similar to the ILI value obtained during the learning process. The main difference between both is that IOC-simple first determines the statistical correlation between call lists from the unknown device and the corresponding GTP class. Here, an assignment criterion is also used to convert calls into specific $\delta_i$ values. 

\begin{equation}\label{eq:Corr}
IOC-simple_{O_{GTP},O_{unk}} = \frac{\sum{o_{GTP}o_{unk}-n{\overline{o_{GTP}}}{\overline{o_{unk}}}}}{ns_{o_{GTP}}s_{o_{unk}}},
\end{equation}
where $o_{GTP}$ represents the set of individual calls in the GTP and $o_{unk}$ represents the set of individual calls extracted from the unknown device. \\

\indent \textit{Index of Correlation Advanced--} As mentioned in Section \ref{sec:Background}, one should not expect smart grid devices to perform operations in a completely deterministic pattern. This limitation exposes the third architectural challenge of our framework (see Section \ref{sec:Background}) since legitimate random operations can be mistaken as compromised behavior. To overcome this constraint, we further apply a more advanced IOC calculation (IOC-advanced). In IOC-advanced, our framework combines the values from $O_{i_unk}$ to $O_{i+h_unk}$ in $O_{unk}$. This operation results in a new random vector $O^\prime_{unk}$ smaller in size and with a lower random component. The index $h$ represents the number of individual calls from the original list that are combined to create the new set $O^\prime_{unk}$. This index value $h$ is proportional to the amount of randomness that one intends to remove from the original $O_{unk}$ and constitutes another configurable parameter in our framework. 

\begin{algorithm}[ht]
\footnotesize
\caption{Steps for the detection and decision processes.}
 \label{algo:architecture}
 \begin{algorithmic}[1]

\STATE $compromised \gets$ 0                                    \label{line:ini1}
\STATE $UNK(DeC_{unk}, SCL_{unk}) \gets$ unknown device profile \label{line:ini2}
\STATE $GTP(DeC_{gtp}, SCL, \sigma) \gets$ GTPs from Database \\         \label{line:ini3}
Detection:
\IF{ Exists $DeC_{gtp} \And DeC_{unk} == DeC_{gtp}$}            \label{line:ifsig}
    \STATE Calculate $IOC$                                      \label{line:IOC}
\ENDIF \\
 Decision:
 \IF{$IOC < \beta$} \label{line:final1}
     \STATE $compromised \gets 1$             \label{line:final2}
\ENDIF
 \end{algorithmic}
 \end{algorithm}

\subsection{Decision Process}
The final step of our framework is the decision process. In this step, our framework compares results from the three detection mechanisms against a threshold $\beta$ to decide if the unknown smart grid device is compromised or not. {\color{black} The value of $\beta$ depends on the device class, and it is always a function of the threshold $\sigma$ determined during the training process and stored in the GTP. The relationship between $\sigma$ and $\beta$ values depends on the targeted accuracy performance for every device class.  In general, for devices with a higher deterministic behavioral pattern, a higher value of $\beta$ is recommended. This design approach reduces the chances of false negatives during the decision process. On the other hand, for devices with lower deterministic behavior, a lower value of $\beta$ may be sufficient to reduce false negatives. Finally, note that this decision threshold is also configurable. The initial value of $\beta$ for every device class can be adjusted to an optimal in real-time and while the framework monitors the devices in the field. In the next section, we analyze practical values of $\beta$ for different types of device classes.}

Finally, Algorithm \ref{algo:architecture} details the detection-decision process of the proposed framework. In lines \ref{line:ini1}, \ref{line:ini2}, and \ref{line:ini3} the variables $compromised$, $UNK$, and $GTP$ are initialized with 0, the profile of the unknown device, and all the signatures from the database, respectively. Then, if a signature of the unknown device's class exists (Line \ref{line:ifsig}), the values $IOC$ (simple and advanced) are calculated in Line \ref{line:IOC}. Finally, if the value of $IOC$ is lower than the threshold $\beta$, the device is decided as \textit{compromised}.


\section{Performance Analysis and Discussion}
\label{sec:performance}
In this section, we analyze the performance of the proposed framework. In all the cases, we obtain the results after averaging 30 different runs of all the covered scenarios. The scenarios include six different types of attackers as a result of the combination of three different threats and two different types of devices based on their computational resources, as described in Section~\ref{sec:Background}. Also, we assume that the devices are correctly grouped based on their type. Moreover, we measure the accuracy of our framework with accuracy, precision, recall, and specificity metrics. Finally, we evaluate the performance of the proposed framework in terms of its overhead (e.g., CPU utilization, memory usage, and execution time).

\subsection{Evaluation methodology with a realistic testbed}
Our framework considers a realistic scenario from a smart grid substation. The testbed's configuration includes a publisher-subscriber two-way communications configuration which sends and receives IEC61850-compliant GOOSE messages \cite{iec61850_1}. For this purpose, we utilize an open-source version of IEC61850~\cite{libiec61850} protocol running on Linux-based systems. Our resource-limited devices (i.e., GOOSE publishers) run on a Raspberry Pi 2B, using Advanced RISC Machine (ARM) 32 bits architecture with limited memory and CPU. On the other hand, the resource-rich devices (i.e., GOOSE subscribers) run on a Linux Ubuntu 14.04 system with a more powerful CPU and higher memory configuration. Finally, we utilize two different hooking techniques: ptrace (that performs function call tracing) and library interposition (that performs system call tracing). In our configuration, the publishers open the communication session and wait for the subscribers to connect. Once the devices create and open the communication sockets, the publishers start sending GOOSE messages to the subscribers every one second for a total time interval $t$ of 60 seconds. After the $t$ seconds, the devices close their communication channels. For every compromised device, the malicious threat is active $n$ times during the communication sessions as described in the adversary model (see Section~\ref{sec:AdversaryModel2}). Finally, as detailed in Table \ref{tab:devices}, compromised devices $CD_{1}$, $CD_{2}$, and $CD_{3}$ correspond to resource-limited devices of any class that have been compromised with Threats 1, 2, and 3 respectively (see Section \ref{sec:AdversaryModel2}) and compromised devices $CD_{4}$, $CD_{5}$, and $CD_{6}$ correspond to resource-rich devices of any class that have been compromised with Threats 1, 2, and 3, respectively. Despite that the initial application of our testbed was intended to evaluate the performance of the proposed framework in realistic scenarios, we believe that, due to its open-source and configurable nature, it can also be used as a benchmark to effectively evaluate the performance of other security tools applied to the smart grid.

\subsection{Detection performance}
In the following, we detail the performance of our framework after applying the three detection mechanisms proposed in Section \ref{sec:Framework}.

\subsubsection{System and Function call lists comparison}
Tables~\ref{tab:server} and~\ref{tab:client} summarize some of the system and function calls captured from the resource-limited and the resource-rich devices, respectively. Columns \textit{Genuine} and $CD_{i}$ (i: 1 to 6) in both tables list the average rate of the system and function calls normalized against the GTP for genuine and compromised devices, respectively. 

\begin{table}[ht!]
\begin{center}
\scriptsize
\caption{Normalized rate of the system and function calls captured after using our framework to detect compromised \textit{resource-limited devices} (e.g., RTUs, PLCs): calls due to malicious activities are grayed.} \vspace{-.25cm}
\label{tab:server}
\begin{tabular}{|c|c|c|c|c|c|} 
\hline
 Call Tracing Technique  & Type of Call & Genuine & $CD_{1}$ & $CD_{2}$ & $CD_{3}$ \\
\hline
\multirow{11}{5em}{ptrace}    
                                 & $brk$ & $\sim$1 & $\sim$1 & \cellcolor{gray!25}6.7 & $\sim$1\\ 
                                 & $clone$ & $\sim$1 & \cellcolor{gray!25}12.5 & $\sim$1 & $\sim$1\\
                                 & $close$ & $\sim$1 & $\sim$1 & $\sim$1 & \cellcolor{gray!25}3.2\\
                                 & $fstat64$ & $\sim$1 & $\sim$1 & $\sim$1 & \cellcolor{gray!25}8.8\\
                                 & $lseek$ & $\sim$1 & $\sim$1 & $\sim$1 & $\sim$1\\
                                 & $mmap2$ & $\sim$1 & \cellcolor{gray!25}2.4 & \cellcolor{gray!25}4.4 & \cellcolor{gray!25}2.4\\
                                 & $mprotect$ & $\sim$1 & \cellcolor{gray!25}2.8 & 1.1 & 1\\
                                 & $munmap$ & $\sim$1 & $\sim$1 & \cellcolor{gray!25}2 & \cellcolor{gray!25}13\\
                                 & $open$ & $\sim$1 & $\sim$1 & $\sim$1 & \cellcolor{gray!25}5\\
                                 & $rt\_sigprocmask$ & $\sim$1 & \cellcolor{gray!25}8.7 & \cellcolor{gray!25}0.3 & \cellcolor{gray!25}0.3\\
                                 & $rt\_sigaction$ & $\sim$1 & $\sim$1 & \cellcolor{gray!25}3 & \cellcolor{gray!25}3\\
\hline
\multirow{12}{5em}{Interposition} & $close$ & $\sim$1 & $\sim$1 & $\sim$1 & $\sim$1\\
                                 & $free$ & $\sim$1 & \cellcolor{gray!25}3.2 & $\sim$1 & $\sim$1\\ 
                                 & $malloc$ & $\sim$1 & \cellcolor{gray!25}3.3 & $\sim$1 & $\sim$1\\ 
                                 & $memcpy$ & $\sim$1 & $\sim$1 & $\sim$1 & $\sim$1\\
                                 & $memset$ & $\sim$1 & $\sim$1 & $\sim$1 & $\sim$1\\
                                 & $mmap$ & $\sim$1 & \cellcolor{gray!25}12.5 & $\sim$1 & $\sim$1\\
                                 & $mprotect$ & $\sim$1 & \cellcolor{gray!25}12.5 & $\sim$1 & $\sim$1\\
                                 & $pthread\_create$ & $\sim$1 & \cellcolor{gray!25}12.5 & $\sim$1 & $\sim$1\\
                                 & $sendto$ & $\sim$1 & \cellcolor{gray!25} 4.3 & $\sim$1 & $\sim$1\\
                                 & $signal$ & $\sim$1 & \cellcolor{gray!25}24 & $\sim$1 & $\sim$1\\
                                 & $socket$ & $\sim$1 & $\sim$1 & $\sim$1 & $\sim$1\\
                                 & $usleep$ & $\sim$1 & \cellcolor{gray!25}3.5 & $\sim$1 & $\sim$1\\
\hline
\end{tabular}
\end{center} \vspace{-0.2cm}
\end{table}

\begin{table}[ht!]
\begin{center}
\scriptsize
\caption{Normalized rate of system and function calls captured after using our framework to detect compromised \textit{resource-rich devices} (e.g., PMUs, IEDs): calls due to malicious activities are grayed.}
\label{tab:client} \vspace{-.15cm}
\begin{tabular}{|c|c|c|c|c|c|} 
\hline
 Call Tracing Technique & Type of Call & Genuine & $CD_{4}$ & $CD_{5}$ & $CD_{6}$ \\
\hline
\multirow{9}{5em}{ptrace}    
                                 & $brk$ & $\sim$1 & $\sim$1 & \cellcolor{gray!25}8.3 & $\sim$1\\ 
                                 & $clone$ & $\sim$1 & \cellcolor{gray!25}23 & $\sim$1 & $\sim$1\\
                                 & $close$ & $\sim$1 & \cellcolor{gray!25}6.5 & \cellcolor{gray!25}6.8 & \cellcolor{gray!25}6.75\\
                                 & $fstat$ & $\sim$1 & \cellcolor{gray!25}12 & \cellcolor{gray!25}12.5 & \cellcolor{gray!25}12.25\\
                                 & $mmap$ & $\sim$1 & \cellcolor{gray!25}4.1 & \cellcolor{gray!25}6.64 & \cellcolor{gray!25}2.6\\
                                 & $mprotect$ & $\sim$1 & \cellcolor{gray!25}3.4 & $\sim$1 & $\sim$1\\
                                 & $munmap$ & $\sim$1 & \cellcolor{gray!25}23 & \cellcolor{gray!25}26 & \cellcolor{gray!25}24\\
                                 & $open$ & $\sim$1 & \cellcolor{gray!25}6.5 & \cellcolor{gray!25}6.75 & \cellcolor{gray!25}6.8\\
                                 & $rt\_sigaction$ & $\sim$1 & \cellcolor{gray!25}8.3 & $\sim$1 & $\sim$1\\
\hline
\multirow{12}{5em}{Interposition} & $free$ & $\sim$1 & \cellcolor{gray!25}15.6 & $\sim$1 & $\sim$1\\ 
                                 & $malloc$ & $\sim$1 & \cellcolor{gray!25}15.6 & $\sim$1 & $\sim$1\\ 
                                 & $memcpy$ & $\sim$1 & \cellcolor{gray!25}17.8 & $\sim$1 & $\sim$1\\
                                 & $memset$ & $\sim$1 & \cellcolor{gray!25}24 & $\sim$1 & $\sim$1\\
                                 & $mmap$ & $\sim$1 & \cellcolor{gray!25}24 & $\sim$1 & $\sim$1\\
                                 & $mprotect$ & $\sim$1 & \cellcolor{gray!25}24 & $\sim$1 & $\sim$1\\
                                 & $pthread\_create$ & $\sim$1 & \cellcolor{gray!25}24 & $\sim$1 & $\sim$1\\
                                 & $pthread_detach$ & $\sim$1 & \cellcolor{gray!25}24 & $\sim$1 & $\sim$1\\
                                 & $recvfrom$ & $\sim$1 & \cellcolor{gray!25}15.7 & $\sim$1 & $\sim$1\\
                                 & $signal$ & $\sim$1 & \cellcolor{gray!25}24 & $\sim$1 & $\sim$1\\
                                 & $socket$ & $\sim$1 & \cellcolor{gray!25}24 & $\sim$1 & $\sim$1\\
                                 & $usleep$ & $\sim$1 & \cellcolor{gray!25}15.7 & $\sim$1 & $\sim$1\\
                                 
\hline
\end{tabular}
\end{center} \vspace{-0.22cm}
\end{table}

\textit{Values greater than $\sim$1 (marked in gray) in columns $CD_{1}$ to $CD_{3}$ and $CD_{4}$ to $CD_{6}$ represents extra system or function call activity due to the presence of malicious operations. That is, extra call activity reveals the presence of malicious activity in the devices.} One can notice that, by using ptrace, our framework identified all cases of compromised devices. On the other hand, in the case of library interposition, only $CD_{1}$ and $CD_{4}$ were properly detected. Also, the reader can notice that in the case of genuine devices, the normalized rate values of system and function calls are very close to 1 in all the cases.

\begin{figure*}[ht]
    \centering
    \subfigure[]{\includegraphics[width=0.5\columnwidth]{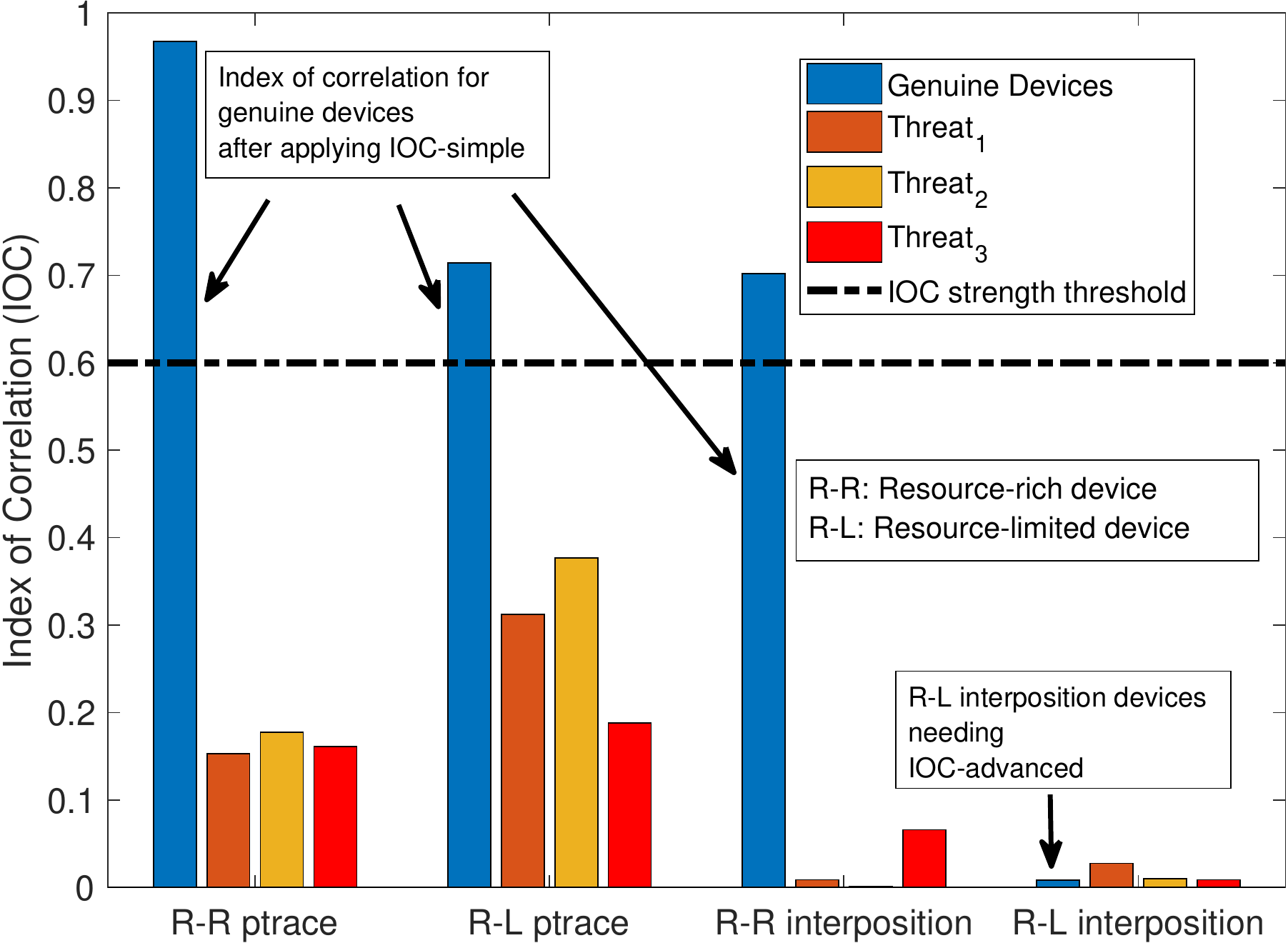}\label{fig:IOCsimple}}
    \subfigure[]{\includegraphics[width=0.475\columnwidth]{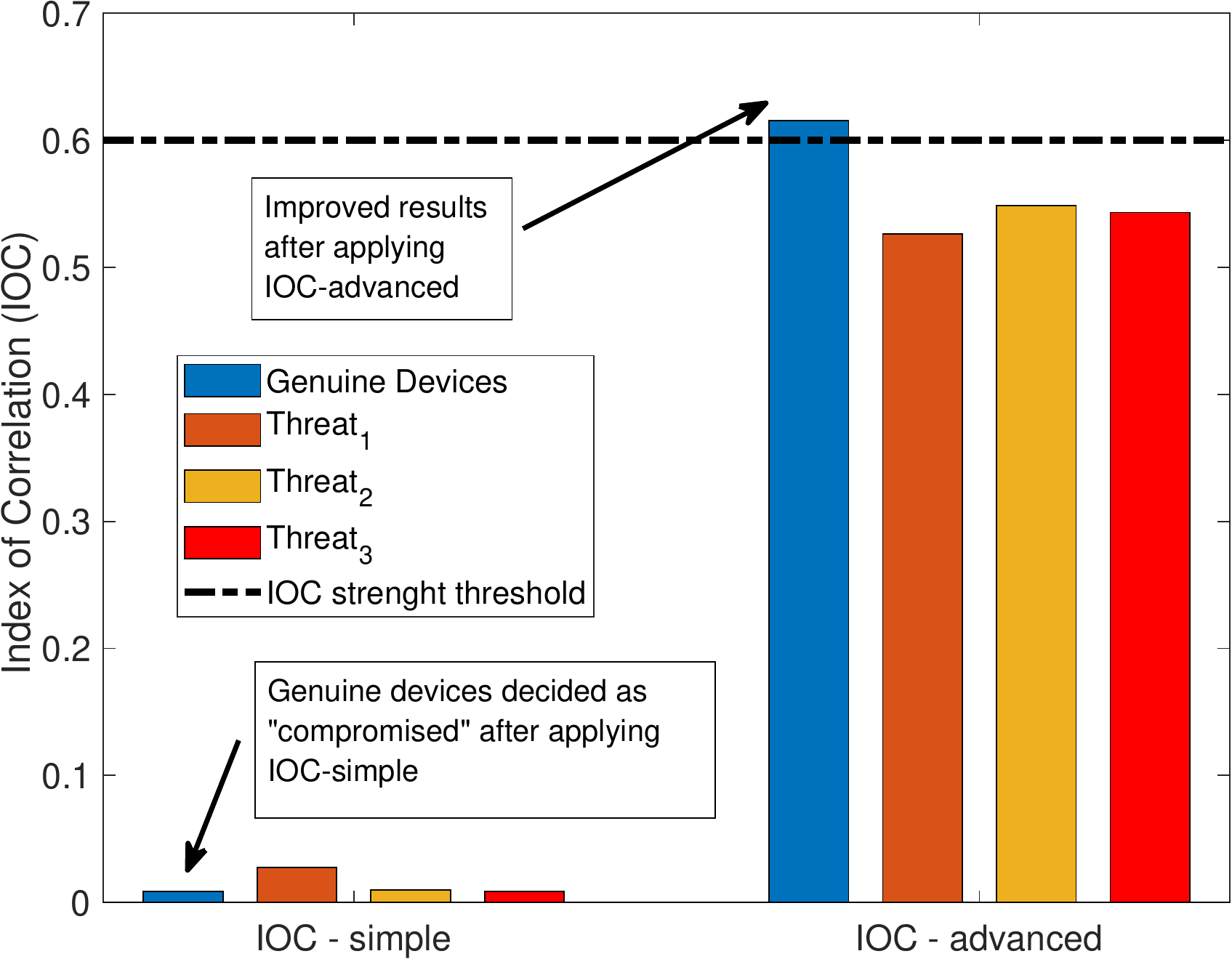}\label{fig:IOCadvanced}}
    \caption{Index of Correlation between GTP and unknown devices: (a) Resource-rich and resource-limited devices after applying our IOC-simple and (b) IOC-advanced results comparison between genuine and compromised resource-limited devices (using  system call lists from library interposition only).}
\end{figure*}

\subsubsection{IOC-simple} The first detection approach could not identify Threats 2 and 3 when the framework utilized library interposition. To overcome this limitation, we applied our second detection mechanism, IOC-simple. As explained in Section \ref{sec:Framework}, to utilize the framework efficiently, the framework applies the different detection approaches in an ordered fashion as needed.

Figure~\ref{fig:IOCsimple} shows the results after applying IOC-simple to system and function call lists from GTP and compromised devices. In this figure, \textit{R-R} refers to resource-rich devices, and \textit{R-L} refers to resource-limited devices.

The reader can observe that, by using ptrace, we obtain low IOC values (in the range of 0.15 to 0.35) between function call lists from GTP and compromised devices. By setting the correlation strength threshold to 0.6 (moderate to high correlation~\cite{RossBook}), our framework detects all the cases of the compromised devices. For the case of library interposition, the framework performs very well for resource-rich compromised devices. However, for resource-limited compromised devices IOC-simple under-performs when the framework applies library interposition. In this particular case, IOC-simple from genuine devices falls under the threshold, triggering false positive results. We relate these results to higher random activity in the resource-limited compromised devices' kernel~\cite{systemcalls5}.

\subsubsection{IOC-advanced} To overcome the previous limitation, we can apply the IOC-advanced technique. By using this approach, our framework can obtain new call lists with more deterministic behavior from the resource-limited devices and enhance the statistical correlation between these type of devices and their corresponding GTP. In Figure~\ref{fig:IOCadvanced}, the reader can observe how IOC values from resource-limited genuine devices overcome the threshold mark while the compromised devices are still under the borderline. There exists a trade-off between the amount of randomness that can be removed from system call lists without impacting the decision process. If the value of $h$ is too significant, critical behavioral information can also be potentially removed from the call lists, limiting the performance of the decision algorithm in cases where tasks $T$ are too simple. 

\subsection{Performance Metrics}
To further measure the efficacy of our detection methods, we calculate the standard performance metrics of accuracy, recall, precision, and specificity. These metrics are defined in Equations~\ref{eq:accuracy}, \ref{eq:recall}, \ref{eq:precision}, and \ref{eq:specificity}:
\begin{equation}\label{eq:accuracy}
A_{CC} = \frac{(T_P+T_N)}{(T_P+T_N+F_P+F_N)},
\end{equation}
\begin{equation}\label{eq:recall}
R_{EC} = \frac{T_P}{(T_P+F_N)},
\end{equation}
\begin{equation}\label{eq:precision}
P_{REC} = \frac{T_P}{(T_P+F_P)},
\end{equation}
\begin{equation}\label{eq:specificity}
S_{pec} = \frac{T_N}{(T_N+F_P)}.
\end{equation}
where $T_P$ stands for true positive or the case where a compromised device is decided as compromised; $T_N$ stands for true negative or the case where a genuine device is decided as genuine; $F_P$ stands for false positive or the case where a genuine device is decided as compromised; and finally $F_N$ stands for false negative or the case where a compromised device is decided as genuine. First, we evaluate the performance of our framework with IOC-simple. Then, the improved results are shown after applying IOC-advanced.

\begin{figure}[ht]
	\centering
	\subfigure[Accuracy]{\includegraphics[width=0.48\textwidth]{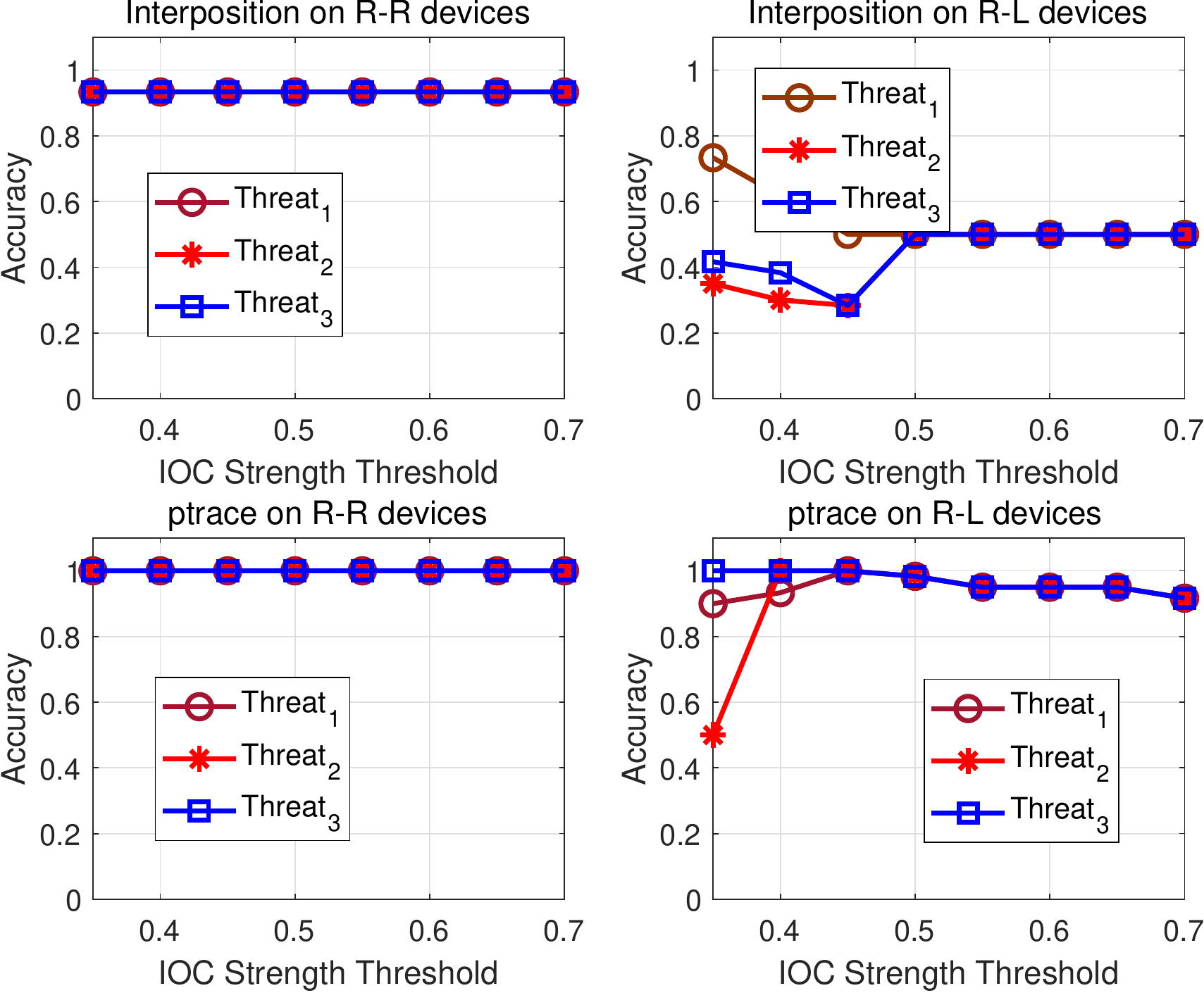}\label{fig:accuracy}}
	\subfigure[Recall]{\includegraphics[width=0.48\textwidth]{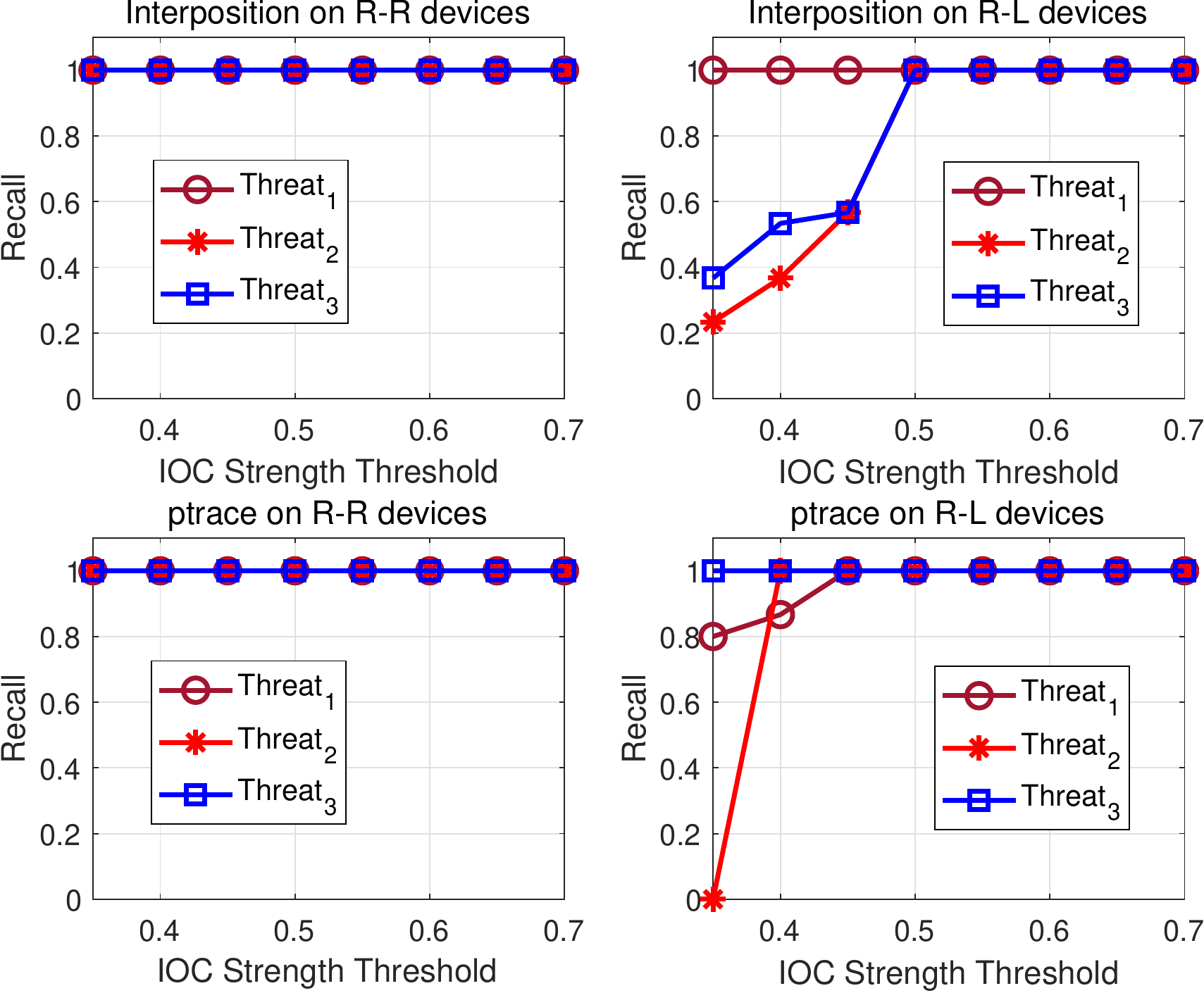}\label{fig:recall}}
	\caption{Figures compare the performance of the {\color{black}IOC-simple algorithm} on six different types of compromised devices after using library interposition and ptrace: (a) Accuracy, (b) Recall.}
	\label{fig:set1}
\end{figure}

In Figure~\ref{fig:accuracy}, we evaluate the overall accuracy of our detection techniques over the six different types of compromised devices. Since accuracy comprises $T_P$ and $T_N$ results, this metric describes how well the framework can positively decide between genuine and compromised devices without errors. In general, for ptrace, our framework achieves an excellent accuracy performance (between 0.95 to 1) for all types of devices. However, this analysis also reveals the performance limitations of the framework for detecting resource-limited compromised devices in the case of library interposition (top-right case in Figure \ref{fig:accuracy}). Here, the framework achieves a low accuracy value of 0.5.

In Figure~\ref{fig:recall}, we evaluate the overall recall performance of the framework. In this case, recall metrics show how well our framework detects the six different types of compromised devices. Based on these results, the reader can observe that the framework achieves the maximum recall (maximum value of $T_P$s) for the selected threshold $\beta$ = 0.6. In the case of resource-rich devices, recall performance was high for all the threshold values. On the other hand, for resource-limited devices, we can notice low recall values for Threats 2 and 3 when the threshold values are under 0.6 for the case of library interposition. 

\begin{figure*}[ht]
	\centering
	\subfigure[Precision]{\includegraphics[width=0.48\textwidth]{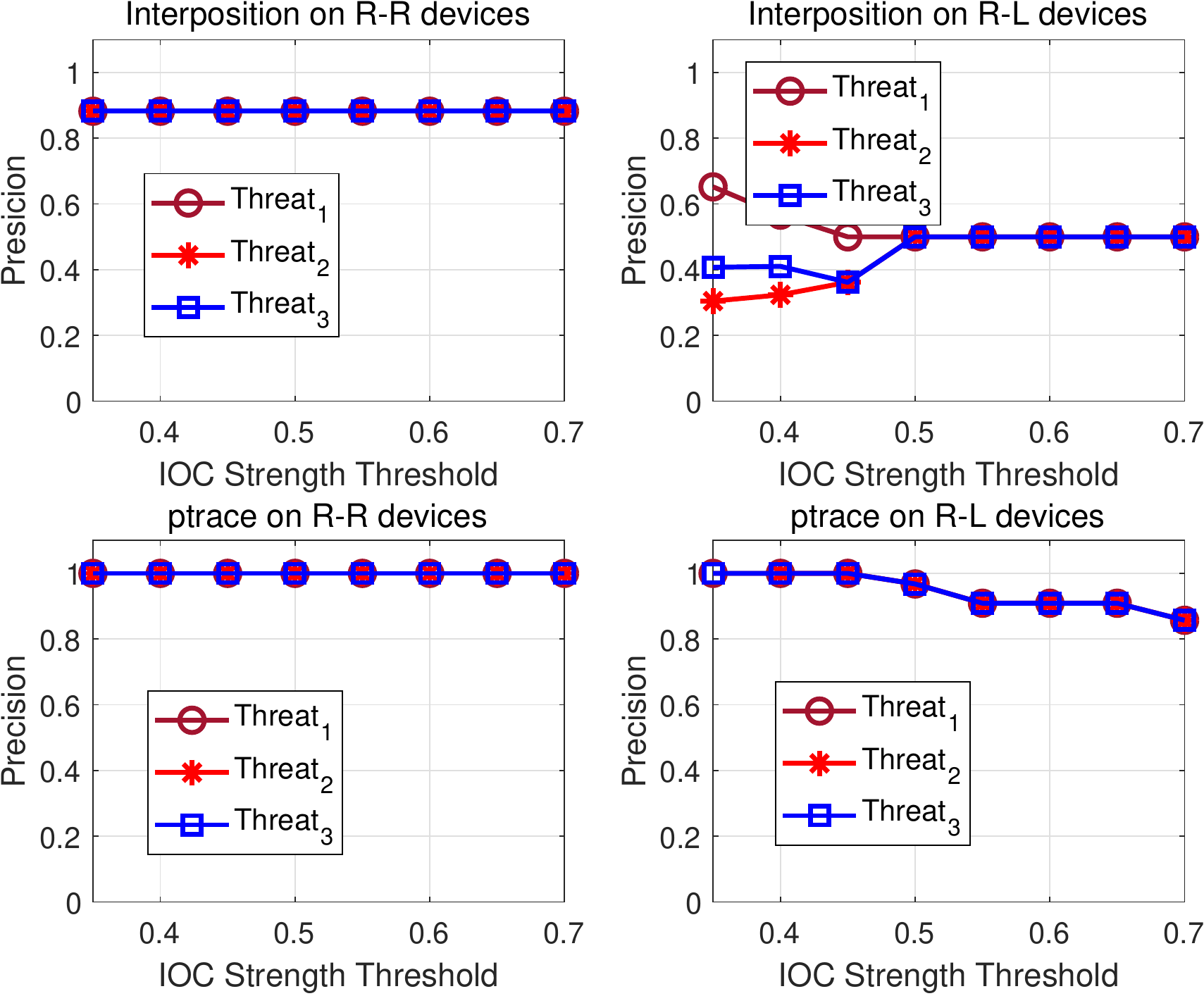}\label{fig:precision}}
	\subfigure[Specificity]{\includegraphics[width=0.48\textwidth]{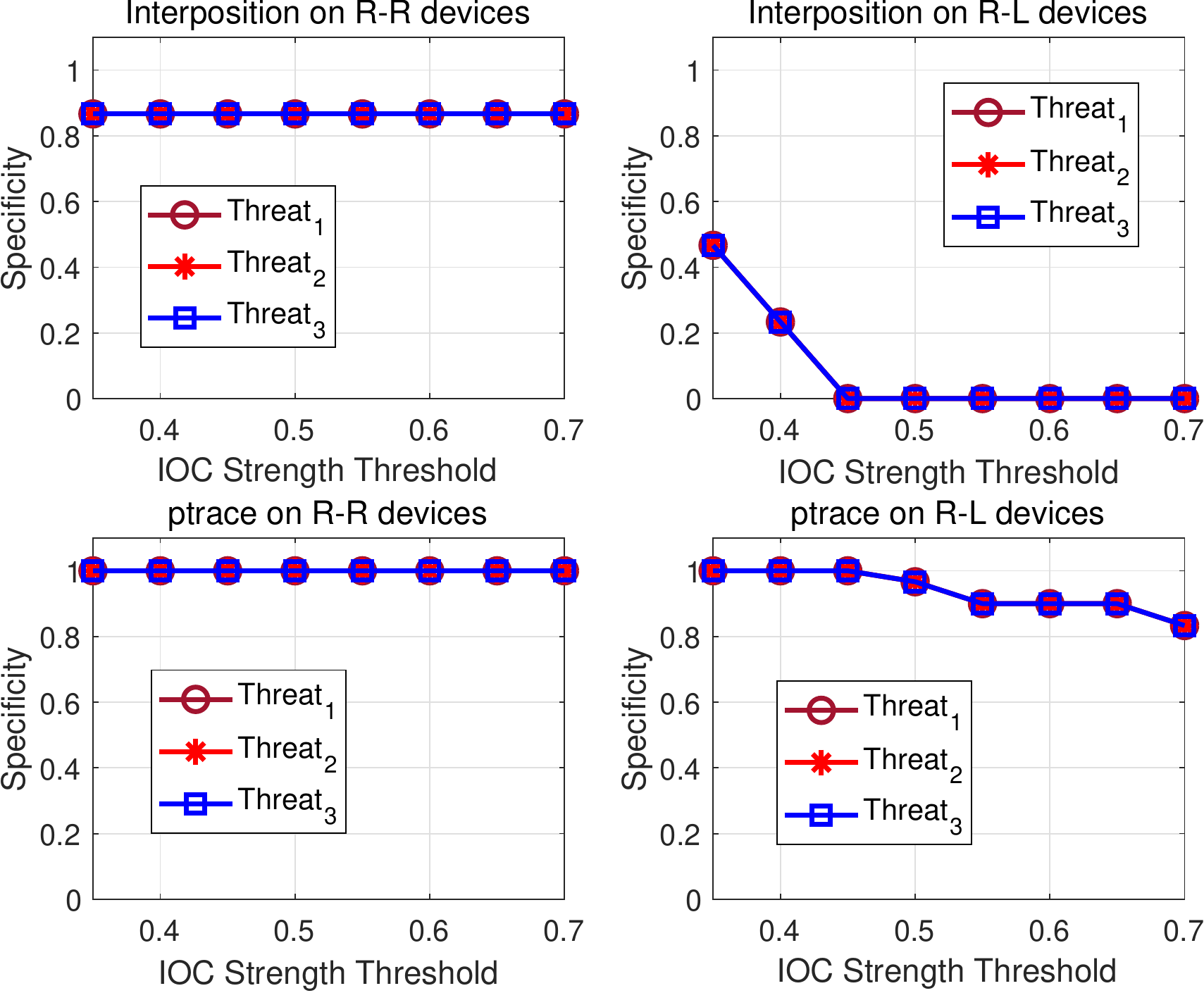}\label{fig:specificity}}
	\caption{Figures compare the performance of the {\color{black}IOC-simple algorithm} on six different types of compromised devices after using library interposition and ptrace: (a) Precision, (b) Specificity.}
	\label{fig:set2}
\end{figure*}

Figure~\ref{fig:precision} depicts the precision evaluation. Precision values represent the statistical relationship between the number of successfully detected compromised devices against the number of times that the framework fails to correctly decide a device as genuine. By looking at the precision results, one can observe that our framework under-performs in the case of library interposition for resource-limited devices. 

Finally, we utilize specificity metrics to evaluate the true negative rate, that is, how effectively our framework discriminates genuine devices. In Figure \ref{fig:specificity} (top right), one can observe that, for the case of resource-limited devices with library interposition, the framework achieves very low specificity. These results limit the application of IOC-simple to decide on this particular type of devices. Specificity value of 0 at $\beta$ threshold between 0.45 and 0.7 demonstrates that a device was not correctly decided as genuine in this case. However, in all the remaining three cases, the framework performs very well.

\begin{figure}[ht]
\vspace{-0.25cm}
    \centering{\includegraphics[width=0.75\textwidth]{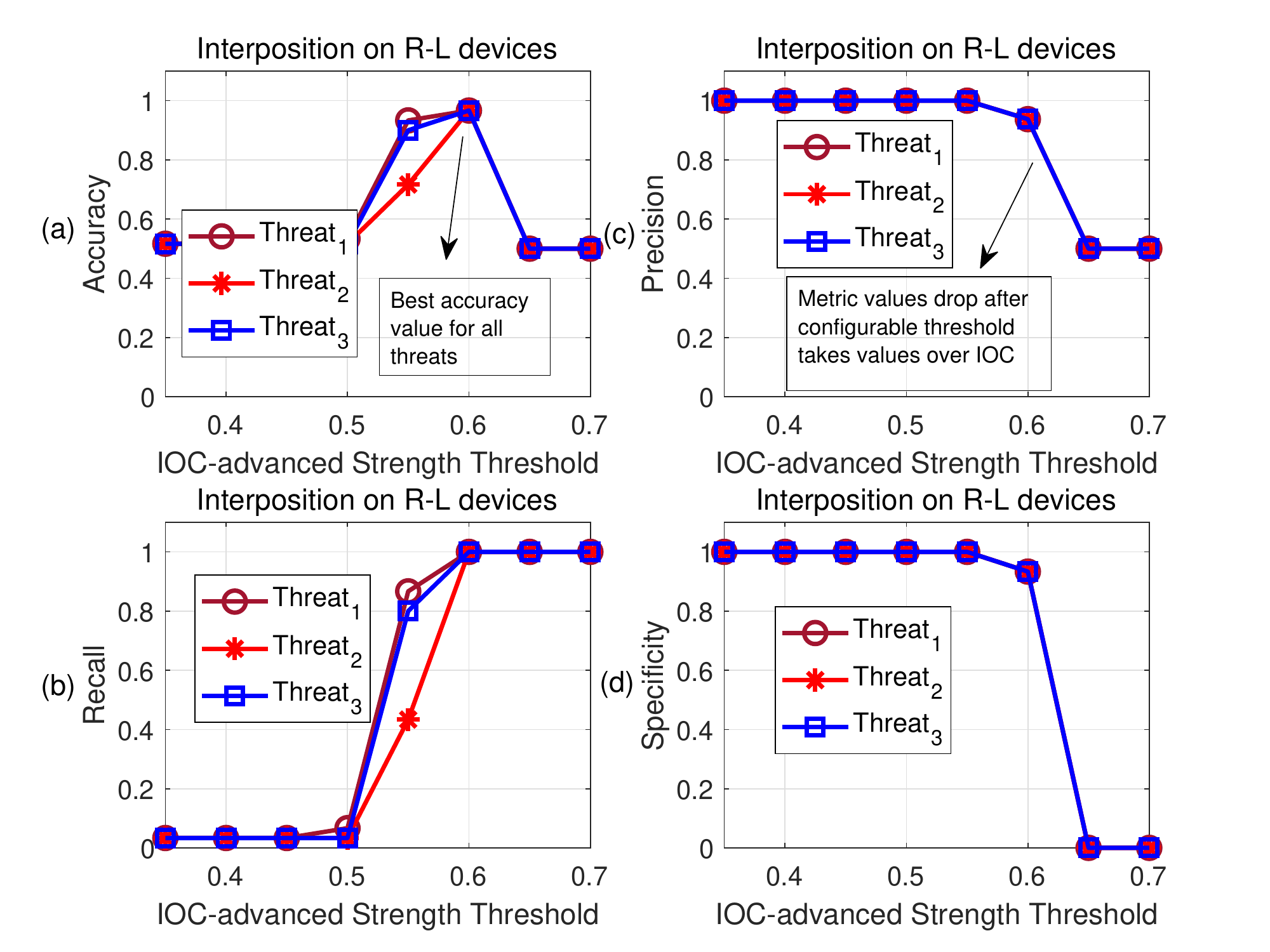}}
    \caption{Performance metrics after applying IOC-advanced for the detection of resource-limited devices when library interposition is utilized: (a) Accuracy, (b) Recall, (c) Precision, and (d) Specificity.}\vspace{-0.25cm}
    \label{fig:All}
\end{figure}

{\color{black}By analyzing the results in Figures \ref{fig:set1} and \ref{fig:set2}, one can compare the performance of the proposed framework on resource-limited and resource-rich devices for the two hooking techniques applied. Most evaluation metrics diminish their performance when the framework applies the IOC-simple algorithm to detect resource-limited devices using library interposition. These results reflect on the fact that for this type of devices, a more robust detection mechanism is necessary. To improve these results, we utilize the framework with the IOC-advanced algorithm. Figure \ref{fig:All} depicts the improvements in all the performance metrics after applying IOC-advanced for compromised resource-limited devices with library interposition. In this figure, one can observe that the correlation threshold of 0.6 provided the best results overall for this particular testbed. Also, the framework obtained significant improvements in accuracy and precision if compared with the case of IOC-simple (accuracy improved from 0.5 to 0.96, and precision improved from 0.5 to 0.93). Finally, recall metrics retained its high performance at the selected threshold value (recall = 1).}

\subsection{System overhead}
\label{performance}
We expect our framework to perform with high accuracy and scalability without introducing too much overhead. Table~\ref{tab:performance} summarizes the average of system overhead on resource-limited and resource-rich devices. The metrics $RT$, $ST$, $UT$, $Mem$, and $CPU$ correspond to the values of real-time, system-time, user-time, memory, and CPU, respectively. In this table, \textit{NF} (No Framework) represents the case where devices were evaluated without applying our the proposed framework, and \textit{WF} (With Framework) represents the cases where we evaluated the performance while applying the framework. Additionally, \textit{LI} represents the cases where we applied library interposition. Finally, \textit{R-R} refers to resource-rich devices, and \textit{R-L} refers to resource-limited devices. Results in Table \ref{tab:performance} demonstrate that the utilization of the detection framework does not introduce significant overhead on the devices. Particularly, in the case of resource-limited devices, the framework utilizes 0.03\% more of memory (out of the total memory available in the devices) and 1.9\% more of the CPU. For resource-rich devices, the framework utilizes 0.001\% more of memory (out of the total memory available on the device) and an almost negligible amount of CPU. In summary, for both resource-limited and resource-rich devices, library interposition introduces the most overhead to the system. However, this overhead is considerably low if compared with similar applications proposed in the literature~\cite{systemcalls1, systemcalls2}. 

\begin{table}[h!]
\vspace{-0.15cm}
\scriptsize
\centering
\caption{Average system overhead on resource-rich and resource-limited devices after using the framework.}
\label{tab:performance} \vspace{-.25cm}
\begin{tabular}{|c|c|c|c|c|c|c|} 
 \hline
 \textit{Metrics} & \multicolumn{2}{c|}{\textit{NF}} & \multicolumn{4}{c|} {\textit{WF}}\\
  & value & value & \multicolumn{2}{c|}{\textit{ptrace (\%)}} & \multicolumn{2}{c|}{\textit{LI (\%)}}  \\
  & \textit{R-R} & \textit{R-L} & \textit{R-R} & \textit{R-L} & \textit{R-R} & \textit{R-L}\\
 \hline
 RT (s) & 60.00 & 60.11 &\cellcolor{gray!25} 0.05 & \cellcolor{gray!25}3.8 & \cellcolor{gray!25}0.01 & \cellcolor{gray!25}0.1 \\ \hline
 ST (s) & 0.49 & 3.60 & \cellcolor{gray!25}8.1 & \cellcolor{gray!25}3.6 & \cellcolor{gray!25}10.2 & \cellcolor{gray!25}5.5\\ \hline
 UT (s) & 0.31 & 0.49 & \cellcolor{gray!25}16.1 & \cellcolor{gray!25}0.31 & \cellcolor{gray!25}6.4 & \cellcolor{gray!25}2.0\\ \hline
 Mem (KB) & 1967.5 & 1827.5 & \cellcolor{gray!25}1.1e-3 & \cellcolor{gray!25}4.3e-5 & \cellcolor{gray!25}3.0e-2 & \cellcolor{gray!25}1.0e-3\\ \hline
 CPU (\%) & 1 & 6.02 & \cellcolor{gray!25}0 & \cellcolor{gray!25}1.9 & \cellcolor{gray!25}0 & \cellcolor{gray!25}1 \\ 
 \hline
 \end{tabular}
\end{table}

To further study the impact of our framework, we analyzed this overhead considering a real resource-limited smart grid device. In Table~\ref{tab:device}, we summarize the main specifications of Remote Terminal Unit RT2020.
\begin{table}[ht!]
\centering
\scriptsize
\caption{Specification values for Remote Terminal Unit RT2020~\cite{rtu}.}
\label{tab:device} \vspace{-.25cm}
 \begin{tabular}{|c|c|} 
 \hline
 \textit{Item} & \textit{Specification Values}\\
 \hline
 Processor & Dual Core ARM A9 667 MHz\\ \hline
 Dynamic Memory (RAM) & 128 MB\\ \hline
 Program Memory (Flash) & 4 MB\\ \hline
 Nonvolatile Memory & 4 Mb\\ \hline
 Real Time Clock Resolution & 1 ms\\ \hline
 Execution Cycle Time & $\leq$ 100 ms\\
 \hline
 \end{tabular}
\end{table}
Looking at Table~\ref{tab:device}, we can conclude that for the worst case of resource utilization (library interposition on a resource-limited device), the increment in execution time because of the use of our framework would only represent up to 2.3 cycle times. Additionally, our framework would only take 0.1\% of the total memory of a real resource-limited smart grid device.

\subsection{Benefits and Features}
There are several benefits associated with the design of our framework:
\begin{enumerate}
    \item {\textit{Excellent detection rate}: the proposed framework demonstrated an excellent rate for the detection of compromised smart grid devices by combining three different detection methods: system and function call comparison, IOC-simple, and IOC-advanced.}
    \item {\textit{Minimum overhead}: the proposed framework does not represent significant overhead on the use of computing resources.}
    \item {\textit{Specific vs. generic solution}: the proposed framework is designed to address the specific problem of compromised smart grid device detection. The adversary and system model proposed in this work follow the security requirements and architecture characteristics of the smart grid. However, the approaches proposed here for the detection of compromised smart grid devices are perfectly suitable for other CPS security domains outside the smart grid domain.}
    \item {\textit{Comprehensive adversary model}: the adversary model used in this work considers both resource-limited and resource-rich compromised devices. Also, it combines three different threats affecting the smart grid.}
    \item {\textit{Compromised device diversity}: Our framework is suitable for a great range of different compromised devices. The design of our system-level framework makes it also suitable for detecting hardware counterfeiting \cite{counterfeit2, counterfeit4, counterfeit6} as observed from the system level. System and function call comparison and statistical techniques are powerful tools capable of detecting changes in hardware and system configuration. This makes our framework an appealing solution to monitor and detect a wide range of different types of compromised devices.}
\end{enumerate}


\section{Related Work}
In this section, we present the related work. There are several works studying security challenges in the smart grid~\cite{securityissues1, securityissues2, compromised1, SEC_grid_challenges}. In general, cyberattacks against smart grid are categorized into four different groups: denial of services (DoS) attacks, malicious data injection attacks, traffic analysis attacks, and high-level application attacks~\cite{SEC_grid_challenges}. In~\cite{SEC_transmission_subsystem}, \cite{SEC_cyber_attacks}, \cite{DOS1}, and \cite{DOS2}, the authors provide several examples of DoS attacks impacting different parts of the smart grid architecture. Most of these attacks are executed from compromised hosts, servers, and devices inside the smart grid.

Malicious data injection attacks are analyzed in~\cite{SEC_transmission_subsystem, datainjection1, datainjection2}. One compelling case is studied in \cite{statestimationattack1}. In this paper, the authors analyze four different types of attacks in the state estimation process and examine the least-effort data injection attack to find the optimal attack vector.

In the case of traffic analysis attacks, authors in \cite{Phasor} describe how an attacker can monitor and intercept the frequency and timing of transmitted messages to deduce information and user's behavior. In~\cite{SEC_grid_home}, high-level application attacks are described as the way an attacker can disrupt the essential functions of a power system (i.e., state estimation and power flow measurement). 

In general, these are all useful studies, but none of them directly covers the threat of compromised devices in the smart grid. Additionally, in cases where the attackers rely on the use and control of compromised smart grid devices to perform the attacks, only one type of smart grid threat was considered at a time. In this work, our adversary model considers a combination of 3 different threats impacting the smart grid combined with 2 different device resource availability.

\indent \textit{Smart grid compromised device detection:} In general, the topic of compromised devices has not been extensively studied in the literature. In most cases, researchers focus on proposing anomaly detection mechanisms \cite{anomalydetection} for different types of attacks in the smart grid \cite{attackdetection1, attackdetection2, attackdetection4, attackdetection5}, without particularizing on the attack sources (e.g., compromised devices). In a few cases, however, the behavior of the smart grid device is considered. In \cite{attackdetection3}, the authors study the minimal number of compromised sensor that can be used to manipulate a given number of smart grid states effectively. Further, they consider the optimal PMU placement to defend against this type of data integrity attacks. Some works have been proposed in other CPS and industrial environments. In \cite{cusum}, the authors propose a vector-valued model-based cumulative sum procedure to identify compromised sensors in CPS. Even though this work achieves promising results in simulation environments, its threat model only considers false data injection attacks. Also, no results are shown on the overhead introduced to the CPS devices, essential to consider suitable security applications for real-time critical infrastructures like the smart grid. In a different approach, integrity measurement and attestation systems have been proposed to evaluate the integrity of applications in CPS and the Internet of Things (IoT) devices~\cite{saint, lkim, maat, iota, daint, aegis}. Also, the authors in \cite{attestation} apply attestation approaches to detect comprised devices in the CPS. In this work, however, they utilize stimulant-response mechanisms to detect compromised devices based on their specific reaction to controlled inputs, which can also be impractical for the smart grid and results can depend on several undesired networks' and physical channels' dynamics. Other relevant works propose similar attestation approaches \cite{attestationSP, attestationHW} to detect attacks in CPS. However, these works focus on building models of the entire CPS network instead of focusing on individual devices, which impacts the overhead and the general performance of the proposed solutions. Finally, most of these works apply to Wireless Sensor Networks (WSN) and are not directly applicable to the smart grid domain. Finally, in more general approaches, some works proposes the use of data collected from devices to detect malicious operations or specific behavior~\cite{iotdots, advertisement}. 

Intelligent, secure packaging, outbound beaconing, and better tracking systems are some of the countermeasures that are being proposed to prevent the introduction of compromised devices in the smart grid supply chain~\cite{SEC_Project, supplychainproblem, supplychainproblem2}. However, skilled attackers could have remote access to legitimate devices (e.g., RTUs, PMUs, and IEDs) outside the supply chain and create opportunities for tampering smart grid devices in the field.  

\indent \textit{Function and system call tracing techniques for security applications:} function and system call tracing techniques constitute a powerful method for regulating and monitoring applications behaviour~\cite{rookit, convolution, patent2}, so they have been largely used in security applications~\cite{interposition_5}. System and function call tracing techniques can be found in applications like intrusion detection and confinement~\cite{interposition_4}, binary detection of OS functions~\cite{interposition_3}, sandboxing~\cite{interposition_2}, and software portable packages~\cite{interposition_1}. Specifically, in \cite{systemcalls4}, the authors use system call tracing to implement intrusion detection systems (IDS). Also, in~\cite{systemcalls2} and \cite{systemcalls3}, the authors proposed anomaly detection mechanism based on information obtained from system calls behavior analysis. In these cases, the implementation of the security tools resulted too heavy in terms of system overhead. One similar application with improved system overhead can be found in \cite{systemcalls1}. In this case, the proposed solution is required to run continuously and serves the purpose of complementing antivirus software.

\noindent \textbf {Difference from existing work:} \textit{Our framework is different from other discussed solutions which, in most cases, focus on specific threats to the smart grid instead of considering multiple types of threats acting on different type of devices (e.g., resource-rich and resource-limited). As discussed, there are also cases where different approaches are used for the detection of compromised devices and/or monitoring application behavior. Only in a few of these cases, the solution is intended to be applied in the smart grid domain. In addition, to succeed, these solutions need to monitor constantly-changing environments like network traffic and computational systems or need to challenge the devices with specific inputs to study their response, which constitutes a limitation in terms of system overhead, resource utilization, and real-time analysis. Differently, our framework has a simpler model and is lightweight in terms of system overhead while providing excellent detection rate of the compromised smart grid devices while they are performing typical real-time CPS operations.} Also, we propose a configurable framework for both the supply chain and the smart grid operation field which is envisioned friendly and adaptive enough to be easily applied either within supply chain testing scenarios and while the devices are performing real-time operations inside the smart grid infrastructure. Finally, our work can also complement the existing security mechanisms in the smart grid domain with its open-source and configurable nature.


\section{Conclusions}
The smart grid vision depends on the secure and reliable two-way communications between smart devices (e.g., IEDs, PLCs, PMUs). Nonetheless, compromised smart grid devices constitute a serious threat to a healthy and secure distribution of data in the grid. In this work, we designed a system-level configurable framework capable of monitoring and detecting compromised smart grid devices. Our framework combines system and function call tracing techniques (i.e., ptrace, library interposition), signal processing, and statistical analysis (basic and advanced) to detect compromised device behavior. To the best of our knowledge, this is the first work that utilizes these techniques in detecting compromised devices in the smart grid. Moreover, we evaluated the performance of our framework on six different types of compromised devices, conforming to realistic smart grid scenarios. Such devices exchanged smart grid GOOSE messages utilizing an open-source version of the IEC61850 protocol suite. Specifically, we analyzed the efficacy of our framework under six different adversarial settings affecting devices with different resource availability. Experimental results demonstrated that our framework successfully detects different types of compromised device behavior in a variety of different environments with high accuracy.  
Also, our performance analysis reveals that the use of the proposed detection framework yield minimal overhead on the smart grid devices' computing resources. 


\section{Acknowledgment}
This material is partially supported by the U.S. Department of Energy under Award Number DE-OE0000779 and by the U.S. National Science Foundation under Award Number NSF-1663051. Any opinions, findings, and conclusions or recommendations expressed in this material are those of the authors and do not necessarily reflect the views of the funding agencies.

\bibliographystyle{ACM-Reference-Format}

\begin{thebibliography}{00}


\ifx \showCODEN    \undefined \def \showCODEN     #1{\unskip}     \fi
\ifx \showDOI      \undefined \def \showDOI       #1{{\tt DOI:}\penalty0{#1}\ }
  \fi
\ifx \showISBNx    \undefined \def \showISBNx     #1{\unskip}     \fi
\ifx \showISBNxiii \undefined \def \showISBNxiii  #1{\unskip}     \fi
\ifx \showISSN     \undefined \def \showISSN      #1{\unskip}     \fi
\ifx \showLCCN     \undefined \def \showLCCN      #1{\unskip}     \fi
\ifx \shownote     \undefined \def \shownote      #1{#1}          \fi
\ifx \showarticletitle \undefined \def \showarticletitle #1{#1}   \fi
\ifx \showURL      \undefined \def \showURL       #1{#1}          \fi
\providecommand\bibfield[2]{#2}
\providecommand\bibinfo[2]{#2}
\providecommand\natexlab[1]{#1}
\providecommand\showeprint[2][]{arXiv:#2}

\bibitem[\protect\citeauthoryear{{A. Kanovsky, P. Spanik and M.
  Frivaldsky}}{{A. Kanovsky, P. Spanik and M. Frivaldsky}}{2015}]%
        {counterfeit6}
\bibfield{author}{\bibinfo{person}{{A. Kanovsky, P. Spanik and M.
  Frivaldsky}}.} \bibinfo{year}{2015}\natexlab{}.
\newblock \showarticletitle{{Detection of electronic counterfeit components}}.
  In \bibinfo{booktitle}{{\em 2015 16th Int. Scientific Conf. on Electric Power
  Engineering (EPE)}}. \bibinfo{publisher}{IEEE}, \bibinfo{address}{Kouty nad
  Desnou}, \bibinfo{pages}{701 -- 705}.
\newblock


\bibitem[\protect\citeauthoryear{{Aksu}, {Babun}, {Conti}, {Tolomei}, and
  {Uluagac}}{{Aksu} et~al\mbox{.}}{2018}]%
        {advertisement}
\bibfield{author}{\bibinfo{person}{H. {Aksu}}, \bibinfo{person}{L. {Babun}},
  \bibinfo{person}{M. {Conti}}, \bibinfo{person}{G. {Tolomei}}, {and}
  \bibinfo{person}{A.~S. {Uluagac}}.} \bibinfo{year}{2018}\natexlab{}.
\newblock \showarticletitle{Advertising in the IoT Era: Vision and Challenges}.
\newblock \bibinfo{journal}{{\em IEEE Communications Magazine\/}}
  \bibinfo{volume}{56}, \bibinfo{number}{11} (\bibinfo{date}{November}
  \bibinfo{year}{2018}), \bibinfo{pages}{138--144}.
\newblock
\showISSN{0163-6804}
\showDOI{%
\url{https://doi.org/10.1109/MCOM.2017.1700871}}


\bibitem[\protect\citeauthoryear{Ansilla, Vasudevan, JayachandraBensam, and
  Anunciya}{Ansilla et~al\mbox{.}}{2015}]%
        {DOS1}
\bibfield{author}{\bibinfo{person}{J.~D. Ansilla}, \bibinfo{person}{N.
  Vasudevan}, \bibinfo{person}{J. JayachandraBensam}, {and}
  \bibinfo{person}{J.~D. Anunciya}.} \bibinfo{year}{2015}\natexlab{}.
\newblock \showarticletitle{Data security in Smart Grid with hardware
  implementation against DoS attacks}. In \bibinfo{booktitle}{{\em 2015
  International Conference on Circuits, Power and Computing Technologies
  [ICCPCT-2015]}}. \bibinfo{pages}{1--7}.
\newblock
\showDOI{%
\url{https://doi.org/10.1109/ICCPCT.2015.7159274}}


\bibitem[\protect\citeauthoryear{Asbj{\o}rnslett}{Asbj{\o}rnslett}{2009}]%
        {supplychainbook}
\bibfield{author}{\bibinfo{person}{Bj{\o}rn~Egil Asbj{\o}rnslett}.}
  \bibinfo{year}{2009}\natexlab{}.
\newblock \bibinfo{booktitle}{{\em Assessing the Vulnerability of Supply
  Chains}}.
\newblock \bibinfo{publisher}{Springer US}, \bibinfo{address}{Boston, MA},
  \bibinfo{pages}{15--33}.
\newblock
\showISBNx{978-0-387-79934-6}
\showDOI{%
\url{https://doi.org/10.1007/978-0-387-79934-6_2}}


\bibitem[\protect\citeauthoryear{Babun, Sikder, Acar, and Uluagac}{Babun
  et~al\mbox{.}}{2018}]%
        {iotdots}
\bibfield{author}{\bibinfo{person}{Leonardo Babun}, \bibinfo{person}{Amit~Kumar
  Sikder}, \bibinfo{person}{Abbas Acar}, {and} \bibinfo{person}{A.~Selcuk
  Uluagac}.} \bibinfo{year}{2018}\natexlab{}.
\newblock \showarticletitle{IoTDots: {A} Digital Forensics Framework for Smart
  Environments}.
\newblock \bibinfo{journal}{{\em CoRR\/}}  \bibinfo{volume}{abs/1809.00745}
  (\bibinfo{year}{2018}).
\newblock
\showeprint[arxiv]{1809.00745}
\showURL{%
\url{http://arxiv.org/abs/1809.00745}}


\bibitem[\protect\citeauthoryear{{Babun, Leonardo (Miami, FL, US), Aksu,
  Hidayet (Miami, FL, US), Uluagac, Selcuk A. (Miami, FL, US)}}{{Babun,
  Leonardo (Miami, FL, US), Aksu, Hidayet (Miami, FL, US), Uluagac, Selcuk A.
  (Miami, FL, US)}}{2018}]%
        {patent}
\bibfield{author}{\bibinfo{person}{{Babun, Leonardo (Miami, FL, US), Aksu,
  Hidayet (Miami, FL, US), Uluagac, Selcuk A. (Miami, FL, US)}}.}
  \bibinfo{year}{2018}\natexlab{}.
\newblock \bibinfo{title}{Detection of counterfeit and compromised devices
  using system and function call tracing techniques}.
\newblock   (\bibinfo{date}{July} \bibinfo{year}{2018}).
\newblock
\showURL{%
\url{http://www.freepatentsonline.com/10027697.html}}


\bibitem[\protect\citeauthoryear{{Babun, Leonardo (Miami, FL, US), Aksu,
  Hidayet (Miami, FL, US), Uluagac, Selcuk A. (Miami, FL, US)}}{{Babun,
  Leonardo (Miami, FL, US), Aksu, Hidayet (Miami, FL, US), Uluagac, Selcuk A.
  (Miami, FL, US)}}{2019}]%
        {patent2}
\bibfield{author}{\bibinfo{person}{{Babun, Leonardo (Miami, FL, US), Aksu,
  Hidayet (Miami, FL, US), Uluagac, Selcuk A. (Miami, FL, US)}}.}
  \bibinfo{year}{2019}\natexlab{}.
\newblock \bibinfo{title}{Method of resource-limited device and device class
  identification using system and function call tracing techniques,
  performance, and statistical analysis}.
\newblock   (\bibinfo{date}{March} \bibinfo{year}{2019}).
\newblock
\showURL{%
\url{http://www.freepatentsonline.com/10242193.html}}


\bibitem[\protect\citeauthoryear{Blunden}{Blunden}{2013}]%
        {rookit}
\bibfield{author}{\bibinfo{person}{Reverend~Bill Blunden}.}
  \bibinfo{year}{2013}\natexlab{}.
\newblock \bibinfo{booktitle}{{\em The Rookit arsenal: Escape and Evasion in
  the Dark Corners of the System\/} (\bibinfo{edition}{2nd} ed.)}.
\newblock \bibinfo{publisher}{Cathleen Sether}, \bibinfo{address}{Burlington,
  MA}.
\newblock
\showISBNx{978-1-4496-2636-5}


\bibitem[\protect\citeauthoryear{Boumkheld, Ghogho, and Koutbi}{Boumkheld
  et~al\mbox{.}}{2016}]%
        {DOS2}
\bibfield{author}{\bibinfo{person}{N. Boumkheld}, \bibinfo{person}{M. Ghogho},
  {and} \bibinfo{person}{M.~El Koutbi}.} \bibinfo{year}{2016}\natexlab{}.
\newblock \showarticletitle{Intrusion detection system for the detection of
  blackhole attacks in a smart grid}. In \bibinfo{booktitle}{{\em 2016 4th
  International Symposium on Computational and Business Intelligence (ISCBI)}}.
  \bibinfo{pages}{108--111}.
\newblock
\showDOI{%
\url{https://doi.org/10.1109/ISCBI.2016.7743267}}


\bibitem[\protect\citeauthoryear{{C. Kriger, S. Behardien and J.
  Retonda-Modiya}}{{C. Kriger, S. Behardien and J. Retonda-Modiya}}{2013}]%
        {iec61850_1}
\bibfield{author}{\bibinfo{person}{{C. Kriger, S. Behardien and J.
  Retonda-Modiya}}.} \bibinfo{year}{2013}\natexlab{}.
\newblock \showarticletitle{{A Detailed Analysis of the GOOSE Message Structure
  in an IEC 61850 Standard-Based Substation Automation System}}.
\newblock \bibinfo{journal}{{\em Int. Journal Comp. Comm.\/}}
  \bibinfo{volume}{8}, \bibinfo{number}{5} (\bibinfo{date}{Oct.}
  \bibinfo{year}{2013}), \bibinfo{pages}{708--721}.
\newblock
\showISSN{1841-9836}


\bibitem[\protect\citeauthoryear{Celik, Babun, Sikder, Aksu, Tan, McDaniel, and
  Uluagac}{Celik et~al\mbox{.}}{2018}]%
        {saint}
\bibfield{author}{\bibinfo{person}{Z.~Berkay Celik}, \bibinfo{person}{Leonardo
  Babun}, \bibinfo{person}{Amit~Kumar Sikder}, \bibinfo{person}{Hidayet Aksu},
  \bibinfo{person}{Gang Tan}, \bibinfo{person}{Patrick McDaniel}, {and}
  \bibinfo{person}{A.~Selcuk Uluagac}.} \bibinfo{year}{2018}\natexlab{}.
\newblock \showarticletitle{Sensitive Information Tracking in Commodity IoT}.
  In \bibinfo{booktitle}{{\em 27th {USENIX} Security Symposium ({USENIX}
  Security 18)}}. \bibinfo{publisher}{{USENIX} Association},
  \bibinfo{address}{Baltimore, MD}, \bibinfo{pages}{1687--1704}.
\newblock
\showISBNx{978-1-931971-46-1}
\showURL{%
\url{https://www.usenix.org/conference/usenixsecurity18/presentation/celik}}


\bibitem[\protect\citeauthoryear{{Ch. Wong and M. Wu}}{{Ch. Wong and M.
  Wu}}{2015}]%
        {counterfeit4}
\bibfield{author}{\bibinfo{person}{{Ch. Wong and M. Wu}}.}
  \bibinfo{year}{2015}\natexlab{}.
\newblock \showarticletitle{{A study on PUF characteristics for counterfeit
  detection}}. In \bibinfo{booktitle}{{\em 2015 IEEE Int. Conf. on Image
  Processing (ICIP)}}. \bibinfo{publisher}{IEEE}, \bibinfo{address}{Quebec
  City, QC}, \bibinfo{pages}{1643 -- 1647}.
\newblock


\bibitem[\protect\citeauthoryear{Chen, Poskitt, and Sun}{Chen
  et~al\mbox{.}}{2018}]%
        {modelling2}
\bibfield{author}{\bibinfo{person}{Y. Chen}, \bibinfo{person}{C.~M. Poskitt},
  {and} \bibinfo{person}{J. Sun}.} \bibinfo{year}{2018}\natexlab{}.
\newblock \showarticletitle{Learning from Mutants: Using Code Mutation to Learn
  and Monitor Invariants of a Cyber-Physical System}. In
  \bibinfo{booktitle}{{\em 2018 IEEE Symposium on Security and Privacy (SP)}}.
  \bibinfo{pages}{648--660}.
\newblock
\showDOI{%
\url{https://doi.org/10.1109/SP.2018.00016}}


\bibitem[\protect\citeauthoryear{{Chen}, {Poskitt}, and {Sun}}{{Chen}
  et~al\mbox{.}}{2018}]%
        {attestationSP}
\bibfield{author}{\bibinfo{person}{Y. {Chen}}, \bibinfo{person}{C.~M.
  {Poskitt}}, {and} \bibinfo{person}{J. {Sun}}.}
  \bibinfo{year}{2018}\natexlab{}.
\newblock \showarticletitle{Learning from Mutants: Using Code Mutation to Learn
  and Monitor Invariants of a Cyber-Physical System}. In
  \bibinfo{booktitle}{{\em 2018 IEEE Symposium on Security and Privacy (SP)}}.
  \bibinfo{pages}{648--660}.
\newblock
\showISSN{2375-1207}
\showDOI{%
\url{https://doi.org/10.1109/SP.2018.00016}}


\bibitem[\protect\citeauthoryear{{Clemens}, {Pal}, and {Sherrell}}{{Clemens}
  et~al\mbox{.}}{2018}]%
        {iota}
\bibfield{author}{\bibinfo{person}{J. {Clemens}}, \bibinfo{person}{R. {Pal}},
  {and} \bibinfo{person}{B. {Sherrell}}.} \bibinfo{year}{2018}\natexlab{}.
\newblock \showarticletitle{Runtime State Verification on Resource-Constrained
  Platforms}. In \bibinfo{booktitle}{{\em MILCOM 2018 - 2018 IEEE Military
  Communications Conference (MILCOM)}}. \bibinfo{pages}{1--6}.
\newblock
\showISSN{2155-7586}
\showDOI{%
\url{https://doi.org/10.1109/MILCOM.2018.8599862}}


\bibitem[\protect\citeauthoryear{Corrigan-Gibbs and Jana}{Corrigan-Gibbs and
  Jana}{2015}]%
        {systemcalls5}
\bibfield{author}{\bibinfo{person}{Henry Corrigan-Gibbs} {and}
  \bibinfo{person}{Suman Jana}.} \bibinfo{year}{2015}\natexlab{}.
\newblock \showarticletitle{Recommendations for Randomness in the Operating
  System or, How to Keep Evil Children out of Your Pool and Other Random
  Facts}. In \bibinfo{booktitle}{{\em Proceedings of the 15th USENIX Conference
  on Hot Topics in Operating Systems}} {\em (\bibinfo{series}{HOTOS'15})}.
  \bibinfo{publisher}{USENIX Association}, \bibinfo{address}{Berkeley, CA,
  USA}, \bibinfo{pages}{25--25}.
\newblock
\showURL{%
\url{http://dl.acm.org/citation.cfm?id=2831090.2831115}}


\bibitem[\protect\citeauthoryear{{D. Formby, P. Srinivasan, A. Leonard, J.
  Rogers, R. Beyah}}{{D. Formby, P. Srinivasan, A. Leonard, J. Rogers, R.
  Beyah}}{2016}]%
        {fingerprinting1}
\bibfield{author}{\bibinfo{person}{{D. Formby, P. Srinivasan, A. Leonard, J.
  Rogers, R. Beyah}}.} \bibinfo{year}{2016}\natexlab{}.
\newblock \showarticletitle{Who\textquotesingle s in Control of Your Control
  System? Device Fingerprinting for Cyber-Physical Systems}. In
  \bibinfo{booktitle}{{\em NDSS, Feb}}.
\newblock


\bibitem[\protect\citeauthoryear{{D. van Opstal, U.S. Resilience Project}}{{D.
  van Opstal, U.S. Resilience Project}}{2012}]%
        {SEC_Project}
\bibfield{author}{\bibinfo{person}{{D. van Opstal, U.S. Resilience Project}}.}
  \bibinfo{year}{2012}\natexlab{}.
\newblock \bibinfo{title}{{Supply chain solutions for smart grid security:
  Building on business best practices.}}
\newblock   (\bibinfo{date}{Sep} \bibinfo{year}{2012}).
\newblock
\showURL{%
\url{http://usresilienceproject.org/wp-content/uploads/2014/09/report-Supply_Chain_Solutions_for_Smart_Grid_Security.pdf}}

\bibitem{aegis}
{\sc Sikder, A.~K., Babun, L., Aksu, H., and Uluagac, A.~S.}
\newblock Aegis: A context-aware security framework for smart home systems.
\newblock {\em arXiv preprint arXiv:1910.03750\/} (2019).

\bibitem{daint}
{\sc Babun, L., Celik Z.~B., McDaniel P., and Uluagac, A.~S.}
\newblock Real-time Analysis of Privacy-(un)aware IoT Applications.
\newblock {\em arXiv preprint arXiv:1911.10461\/} (2019).

\bibitem[\protect\citeauthoryear{Deng and Shukla}{Deng and Shukla}{2012}]%
        {SEC_transmission_subsystem}
\bibfield{author}{\bibinfo{person}{Y. Deng} {and} \bibinfo{person}{S. Shukla}.}
  \bibinfo{year}{2012}\natexlab{}.
\newblock \showarticletitle{{Vulnerabilities and countermeasures - A survey on
  the cyber security issues in the transmission subsystem of a smart grid}}.
\newblock \bibinfo{journal}{{\em Journal of Cyber Security and Mobility\/}}
  \bibinfo{volume}{1} (\bibinfo{year}{2012}), \bibinfo{pages}{251--276}.
\newblock
Issue 4.


\bibitem[\protect\citeauthoryear{{E. Eskin, W. Lee and S. J, Stolfo}}{{E.
  Eskin, W. Lee and S. J, Stolfo}}{2001}]%
        {systemcalls4}
\bibfield{author}{\bibinfo{person}{{E. Eskin, W. Lee and S. J, Stolfo}}.}
  \bibinfo{year}{2001}\natexlab{}.
\newblock \showarticletitle{{Modeling System Calls for Intrusion Detection with
  Dynamic Window Sizes}}. In \bibinfo{booktitle}{{\em DARPA Information
  Survivability Conference \& Exposition II, 2001. DISCEX '01}}.
  \bibinfo{publisher}{IEEE}, \bibinfo{address}{Anaheim, CA},
  \bibinfo{pages}{165--171}.
\newblock


\bibitem[\protect\citeauthoryear{{European Network and Information Security
  Agency (enisa)}}{{European Network and Information Security Agency
  (enisa)}}{2012}]%
        {ENISA}
\bibfield{author}{\bibinfo{person}{{European Network and Information Security
  Agency (enisa)}}.} \bibinfo{year}{2012}\natexlab{}.
\newblock \bibinfo{title}{{Smart Grid Security. Annex II: Security Aspects of
  the Smart Grid.}}
\newblock   (\bibinfo{year}{2012}).
\newblock
\showURL{%
\url{https://www.enisa.europa.eu/topics/critical-information-infrastructures-and-services/smart-grids/smart-grids-and-smart-metering/ENISA_Annex\%20II\%20-\%20Security\%20Aspects\%20of\%20Smart\%20Grid.pdf}}


\bibitem[\protect\citeauthoryear{Farraj, Hammad, Daoud, and Kundur}{Farraj
  et~al\mbox{.}}{2016}]%
        {attackdetection4}
\bibfield{author}{\bibinfo{person}{A. Farraj}, \bibinfo{person}{E. Hammad},
  \bibinfo{person}{A.~A. Daoud}, {and} \bibinfo{person}{D. Kundur}.}
  \bibinfo{year}{2016}\natexlab{}.
\newblock \showarticletitle{A Game-Theoretic Analysis of Cyber Switching
  Attacks and Mitigation in Smart Grid Systems}.
\newblock \bibinfo{journal}{{\em IEEE Transactions on Smart Grid\/}}
  \bibinfo{volume}{7}, \bibinfo{number}{4} (\bibinfo{date}{July}
  \bibinfo{year}{2016}), \bibinfo{pages}{1846--1855}.
\newblock
\showISSN{1949-3053}
\showDOI{%
\url{https://doi.org/10.1109/TSG.2015.2440095}}


\bibitem[\protect\citeauthoryear{Feng, Kolesnikov, Fogla, Lee, and Gong}{Feng
  et~al\mbox{.}}{2003}]%
        {systemcalls2}
\bibfield{author}{\bibinfo{person}{Henry~Hanping Feng},
  \bibinfo{person}{Oleg~M. Kolesnikov}, \bibinfo{person}{Prahlad Fogla},
  \bibinfo{person}{Wenke Lee}, {and} \bibinfo{person}{Weibo Gong}.}
  \bibinfo{year}{2003}\natexlab{}.
\newblock \showarticletitle{Anomaly Detection Using Call Stack Information}. In
  \bibinfo{booktitle}{{\em Proceedings of the 2003 IEEE Symposium on Security
  and Privacy}} {\em (\bibinfo{series}{SP '03})}. \bibinfo{publisher}{IEEE
  Computer Society}, \bibinfo{address}{Washington, DC, USA},
  \bibinfo{pages}{62--}.
\newblock
\showISBNx{0-7695-1940-7}
\showURL{%
\url{http://dl.acm.org/citation.cfm?id=829515.830554}}


\bibitem[\protect\citeauthoryear{Garfinkel}{Garfinkel}{2003}]%
        {interposition_5}
\bibfield{author}{\bibinfo{person}{Tal Garfinkel}.}
  \bibinfo{year}{2003}\natexlab{}.
\newblock \showarticletitle{Traps and Pitfalls: Practical Problems in System
  Call Interposition Based Security Tools}. In \bibinfo{booktitle}{{\em In
  Proc. Network and Distributed Systems Security Symposium}}.
  \bibinfo{pages}{163--176}.
\newblock


\bibitem[\protect\citeauthoryear{Guin, Forte, and Tehranipoor}{Guin
  et~al\mbox{.}}{2013}]%
        {counterfeit2}
\bibfield{author}{\bibinfo{person}{Ujjwal Guin}, \bibinfo{person}{Domenic
  Forte}, {and} \bibinfo{person}{Mohammad Tehranipoor}.}
  \bibinfo{year}{2013}\natexlab{}.
\newblock \showarticletitle{Anti-counterfeit Techniques: From Design to
  Resign}. In \bibinfo{booktitle}{{\em Proceedings of the 2013 14th
  International Workshop on Microprocessor Test and Verification}}.
  \bibinfo{publisher}{IEEE Computer Society}, \bibinfo{address}{Washington, DC,
  USA}, \bibinfo{pages}{89--94}.
\newblock
\showISBNx{978-1-4799-3246-7}
\showURL{%
\url{http://dx.doi.org/10.1109/MTV.2013.28}}


\bibitem[\protect\citeauthoryear{Guo and Engler}{Guo and Engler}{2011}]%
        {interposition_1}
\bibfield{author}{\bibinfo{person}{P.~J. Guo} {and} \bibinfo{person}{D.
  Engler}.} \bibinfo{year}{2011}\natexlab{}.
\newblock \showarticletitle{CDE: Using System Call Interposition to
  Automatically Create Portable Software Packages}. In \bibinfo{booktitle}{{\em
  Proceedings of the 2011 USENIX Conference on USENIX Annual Technical
  Conference}} {\em (\bibinfo{series}{USENIXATC'11})}.
  \bibinfo{publisher}{USENIX Association}, \bibinfo{address}{Berkeley, CA,
  USA}, \bibinfo{pages}{21--21}.
\newblock
\showURL{%
\url{http://dl.acm.org/citation.cfm?id=2002181.2002202}}


\bibitem[\protect\citeauthoryear{Hansen, Staggs, and Shenoi}{Hansen
  et~al\mbox{.}}{2017}]%
        {attackexamples}
\bibfield{author}{\bibinfo{person}{Aaron Hansen}, \bibinfo{person}{Jason
  Staggs}, {and} \bibinfo{person}{Sujeet Shenoi}.}
  \bibinfo{year}{2017}\natexlab{}.
\newblock \showarticletitle{{Security analysis of an advanced metering
  infrastructure}}.
\newblock \bibinfo{journal}{{\em {International Journal of Critical
  Infrastructure Protection}\/}}  \bibinfo{volume}{18} (\bibinfo{year}{2017}),
  \bibinfo{pages}{3 -- 19}.
\newblock
\showISSN{1874-5482}
\showURL{%
\url{http://www.sciencedirect.com/science/article/pii/S1874548217300495}}


\bibitem[\protect\citeauthoryear{Hao, Piechocki, Kaleshi, Chin, and Fan}{Hao
  et~al\mbox{.}}{2014}]%
        {attackdetection2}
\bibfield{author}{\bibinfo{person}{J. Hao}, \bibinfo{person}{R.~J. Piechocki},
  \bibinfo{person}{D. Kaleshi}, \bibinfo{person}{W.~H. Chin}, {and}
  \bibinfo{person}{Z. Fan}.} \bibinfo{year}{2014}\natexlab{}.
\newblock \showarticletitle{Optimal malicious attack construction and robust
  detection in Smart Grid cyber security analysis}. In \bibinfo{booktitle}{{\em
  2014 IEEE International Conference on Smart Grid Communications
  (SmartGridComm)}}. \bibinfo{pages}{836--841}.
\newblock
\showDOI{%
\url{https://doi.org/10.1109/SmartGridComm.2014.7007752}}


\bibitem[\protect\citeauthoryear{He, Chan, and Guizani}{He
  et~al\mbox{.}}{2017a}]%
        {wireless}
\bibfield{author}{\bibinfo{person}{D. He}, \bibinfo{person}{S. Chan}, {and}
  \bibinfo{person}{M. Guizani}.} \bibinfo{year}{2017}\natexlab{a}.
\newblock \showarticletitle{Cyber Security Analysis and Protection of Wireless
  Sensor Networks for Smart Grid Monitoring}.
\newblock \bibinfo{journal}{{\em IEEE Wireless Communications\/}}
  \bibinfo{volume}{24}, \bibinfo{number}{6} (\bibinfo{date}{Dec}
  \bibinfo{year}{2017}), \bibinfo{pages}{98--103}.
\newblock
\showISSN{1536-1284}
\showDOI{%
\url{https://doi.org/10.1109/MWC.2017.1600283WC}}


\bibitem[\protect\citeauthoryear{He, Chan, and Guizani}{He
  et~al\mbox{.}}{2017b}]%
        {win}
\bibfield{author}{\bibinfo{person}{D. He}, \bibinfo{person}{S. Chan}, {and}
  \bibinfo{person}{M. Guizani}.} \bibinfo{year}{2017}\natexlab{b}.
\newblock \showarticletitle{Win-Win Security Approaches for Smart Grid
  Communications Networks}.
\newblock \bibinfo{journal}{{\em IEEE Network\/}} \bibinfo{volume}{31},
  \bibinfo{number}{6} (\bibinfo{date}{November} \bibinfo{year}{2017}),
  \bibinfo{pages}{122--128}.
\newblock
\showISSN{0890-8044}
\showDOI{%
\url{https://doi.org/10.1109/MNET.2017.1700065}}


\bibitem[\protect\citeauthoryear{{Honeywell}}{{Honeywell}}{2014}]%
        {rtu}
\bibfield{author}{\bibinfo{person}{{Honeywell}}.}
  \bibinfo{year}{2014}\natexlab{}.
\newblock \bibinfo{title}{{RTU2020 Remote Terminal Unit Specifications}}.
\newblock   (\bibinfo{date}{Oct} \bibinfo{year}{2014}).
\newblock
\showURL{%
\url{https://www.honeywellprocess.com/library/marketing/tech-specs/SC03-300-101-RTU-2020.pdf}}


\bibitem[\protect\citeauthoryear{Hunt and Brubacher}{Hunt and
  Brubacher}{1999}]%
        {interposition_3}
\bibfield{author}{\bibinfo{person}{G. Hunt} {and} \bibinfo{person}{D.
  Brubacher}.} \bibinfo{year}{1999}\natexlab{}.
\newblock \showarticletitle{Detours: Binary Interception of Win32 Functions}.
  In \bibinfo{booktitle}{{\em Proceedings of the 3rd Conference on USENIX
  Windows NT Symposium - Volume 3}} {\em (\bibinfo{series}{WINSYM'99})}.
  \bibinfo{publisher}{USENIX Association}, \bibinfo{address}{Berkeley, CA,
  USA}, \bibinfo{pages}{14--14}.
\newblock
\showURL{%
\url{http://dl.acm.org/citation.cfm?id=1268427.1268441}}


\bibitem[\protect\citeauthoryear{{IEC 61850-1}}{{IEC 61850-1}}{2003}]%
        {iec61850_4}
\bibfield{author}{\bibinfo{person}{{IEC 61850-1}}.}
  \bibinfo{year}{2003}\natexlab{}.
\newblock \bibinfo{title}{{Communication networks and systems in substations
  Introduction and overview}}.
\newblock   (\bibinfo{year}{2003}).
\newblock
\showURL{%
\url{https://webstore.iec.ch/p-preview/info_iec61850-1\%7Bed1.0\%7Den.pdf}}


\bibitem[\protect\citeauthoryear{{IEC 61850-7-2}}{{IEC 61850-7-2}}{2003}]%
        {iec61850_5}
\bibfield{author}{\bibinfo{person}{{IEC 61850-7-2}}.}
  \bibinfo{year}{2003}\natexlab{}.
\newblock \bibinfo{title}{{ Communication networks and systems in substations -
  Basic communication structure for substation and feeder equipment Abstract
  Communication Service Interface (ACSI)}}.
\newblock   (\bibinfo{year}{2003}).
\newblock
\showURL{%
\url{https://webstore.iec.ch/p-preview/info_iec61850-7-2\%7Bed1.0\%7Den.pdf}}


\bibitem[\protect\citeauthoryear{{IEC 61850-8-1}}{{IEC 61850-8-1}}{2003}]%
        {iec61850_3}
\bibfield{author}{\bibinfo{person}{{IEC 61850-8-1}}.}
  \bibinfo{year}{2003}\natexlab{}.
\newblock \bibinfo{title}{{Communication networks and systems in substations -
  Specific Communication Service Mapping (SCSM) Mappings to MMS (ISO 9506-1 and
  ISO 9506-2) and to ISO/IEC 8802-3}}.
\newblock   (\bibinfo{year}{2003}).
\newblock
\showURL{%
\url{https://webstore.iec.ch/p-preview/info_iec61850-8-1\%7Bed1.0\%7Den.pdf}}


\bibitem[\protect\citeauthoryear{{IEC61850-7-1}}{{IEC61850-7-1}}{2003}]%
        {iec61850_2}
\bibfield{author}{\bibinfo{person}{{IEC61850-7-1}}.}
  \bibinfo{year}{2003}\natexlab{}.
\newblock \bibinfo{title}{{Communication networks and systems for power utility
  automation - Part 7-1: Basic communication structure - Principles and
  models}}.
\newblock   (\bibinfo{year}{2003}).
\newblock
\showURL{%
\url{https://webstore.iec.ch/publication/6014}}


\bibitem[\protect\citeauthoryear{Ingram, Schaub, Taylor, and Campbell}{Ingram
  et~al\mbox{.}}{2013}]%
        {tii}
\bibfield{author}{\bibinfo{person}{D.~M.~E. Ingram}, \bibinfo{person}{P.
  Schaub}, \bibinfo{person}{R.~R. Taylor}, {and} \bibinfo{person}{D.~A.
  Campbell}.} \bibinfo{year}{2013}\natexlab{}.
\newblock \showarticletitle{Performance Analysis of IEC 61850 Sampled Value
  Process Bus Networks}.
\newblock \bibinfo{journal}{{\em IEEE Transactions on Industrial
  Informatics\/}} \bibinfo{volume}{9}, \bibinfo{number}{3} (\bibinfo{date}{Aug}
  \bibinfo{year}{2013}), \bibinfo{pages}{1445--1454}.
\newblock
\showISSN{1551-3203}
\showDOI{%
\url{https://doi.org/10.1109/TII.2012.2228874}}


\bibitem[\protect\citeauthoryear{{Interos Solutions, Inc}}{{Interos Solutions,
  Inc}}{2018}]%
        {supplytransparency}
\bibfield{author}{\bibinfo{person}{{Interos Solutions, Inc}}.}
  \bibinfo{year}{2018}\natexlab{}.
\newblock \bibinfo{title}{Supply Chain Vulnerabilities from China in U.S.
  Federal Information and Communications Technology}.
\newblock   (\bibinfo{year}{2018}).
\newblock
\showURL{%
\url{https://www.uscc.gov/sites/default/files/Research/Interos_Supply\%20Chain\%20Vulnerabilities\%20from\%20China\%20in\%20U.S.\%20Federal\%20ICT_final.pdf}}


\bibitem[\protect\citeauthoryear{{J. Ellperin and A. Entous}}{{J. Ellperin and
  A. Entous}}{2016}]%
        {russianhacking}
\bibfield{author}{\bibinfo{person}{{J. Ellperin and A. Entous}}.}
  \bibinfo{year}{2016}\natexlab{}.
\newblock \bibinfo{title}{Russian operation hacked a Vermont utility, showing
  risk to U.S. electrical grid security, officials say}.
\newblock   (\bibinfo{year}{2016}).
\newblock
\showURL{%
\url{https://www.washingtonpost.com/world/national-security/russian-hackers-penetrated-us-electricity-grid-through-a-utility-in-vermont/2016/12/30/8fc90cc4-ceec-11e6-b8a2-8c2a61b0436f_story.html?utm_term=.7445133366ca}}


\bibitem[\protect\citeauthoryear{Jafary, Repo, {Sepp\"{a}l\"{a}}, and
  Koivisto}{Jafary et~al\mbox{.}}{2017}]%
        {icit}
\bibfield{author}{\bibinfo{person}{P. Jafary}, \bibinfo{person}{S. Repo},
  \bibinfo{person}{J. {Sepp\"{a}l\"{a}}}, {and} \bibinfo{person}{H. Koivisto}.}
  \bibinfo{year}{2017}\natexlab{}.
\newblock \showarticletitle{Security and reliability analysis of a use case in
  smart grid substation automation systems}. In \bibinfo{booktitle}{{\em 2017
  IEEE International Conference on Industrial Technology (ICIT)}}.
  \bibinfo{pages}{615--620}.
\newblock
\showDOI{%
\url{https://doi.org/10.1109/ICIT.2017.7915429}}


\bibitem[\protect\citeauthoryear{Jain and Sekar}{Jain and Sekar}{1999}]%
        {interposition_4}
\bibfield{author}{\bibinfo{person}{K. Jain} {and} \bibinfo{person}{R. Sekar}.}
  \bibinfo{year}{1999}\natexlab{}.
\newblock \showarticletitle{User-Level Infrastructure for System Call
  Interposition: A Platform for Intrusion Detection and Confinement}. In
  \bibinfo{booktitle}{{\em In Proc. Network and Distributed Systems Security
  Symposium}}.
\newblock


\bibitem[\protect\citeauthoryear{{K. Huang, J. M. Carulli, and Y. Makris}}{{K.
  Huang, J. M. Carulli, and Y. Makris}}{2013}]%
        {counterfeithard}
\bibfield{author}{\bibinfo{person}{{K. Huang, J. M. Carulli, and Y. Makris}}.}
  \bibinfo{year}{2013}\natexlab{}.
\newblock \showarticletitle{{Counterfeit electronics: A rising threat in the
  semiconductor manufacturing industry}}. In \bibinfo{booktitle}{{\em ITC. IEEE
  Computer Society}}. \bibinfo{publisher}{IEEE}, \bibinfo{pages}{1--4}.
\newblock


\bibitem[\protect\citeauthoryear{Kang, Adepu, Jackson, and Mathur}{Kang
  et~al\mbox{.}}{2016}]%
        {modelling3}
\bibfield{author}{\bibinfo{person}{E. Kang}, \bibinfo{person}{S. Adepu},
  \bibinfo{person}{D. Jackson}, {and} \bibinfo{person}{A.~P. Mathur}.}
  \bibinfo{year}{2016}\natexlab{}.
\newblock \showarticletitle{Model-Based Security Analysis of a Water Treatment
  System}. In \bibinfo{booktitle}{{\em 2016 IEEE/ACM 2nd International Workshop
  on Software Engineering for Smart Cyber-Physical Systems (SEsCPS)}}.
  \bibinfo{pages}{22--28}.
\newblock
\showDOI{%
\url{https://doi.org/10.1109/SEsCPS.2016.012}}


\bibitem[\protect\citeauthoryear{{Kaspersky}}{{Kaspersky}}{2016}]%
        {black}
\bibfield{author}{\bibinfo{person}{{Kaspersky}}.}
  \bibinfo{year}{2016}\natexlab{}.
\newblock \bibinfo{title}{BlackEnergy APT Attacks in Ukraine}.
\newblock   (\bibinfo{year}{2016}).
\newblock
\showURL{%
\url{https://usa.kaspersky.com/resource-center/threats/blackenergy}}


\bibitem[\protect\citeauthoryear{{Kaygusuz}, {Babun}, {Aksu}, and
  {Uluagac}}{{Kaygusuz} et~al\mbox{.}}{2018}]%
        {convolution}
\bibfield{author}{\bibinfo{person}{C. {Kaygusuz}}, \bibinfo{person}{L.
  {Babun}}, \bibinfo{person}{H. {Aksu}}, {and} \bibinfo{person}{A.~S.
  {Uluagac}}.} \bibinfo{year}{2018}\natexlab{}.
\newblock \showarticletitle{Detection of Compromised Smart Grid Devices with
  Machine Learning and Convolution Techniques}. In \bibinfo{booktitle}{{\em
  2018 IEEE International Conference on Communications (ICC)}}.
  \bibinfo{pages}{1--6}.
\newblock
\showISSN{1938-1883}
\showDOI{%
\url{https://doi.org/10.1109/ICC.2018.8423022}}


\bibitem[\protect\citeauthoryear{Khanna, Panigrahi, and Joshi}{Khanna
  et~al\mbox{.}}{2016}]%
        {datainjection2}
\bibfield{author}{\bibinfo{person}{K. Khanna}, \bibinfo{person}{B.~K.
  Panigrahi}, {and} \bibinfo{person}{A. Joshi}.}
  \bibinfo{year}{2016}\natexlab{}.
\newblock \showarticletitle{Feasibility and mitigation of false data injection
  attacks in smart grid}. In \bibinfo{booktitle}{{\em 2016 IEEE 6th
  International Conference on Power Systems (ICPS)}}. \bibinfo{pages}{1--6}.
\newblock
\showDOI{%
\url{https://doi.org/10.1109/ICPES.2016.7584204}}


\bibitem[\protect\citeauthoryear{Khurana, Hadley, Lu, and Frincke}{Khurana
  et~al\mbox{.}}{2010}]%
        {securityissues1}
\bibfield{author}{\bibinfo{person}{H. Khurana}, \bibinfo{person}{M. Hadley},
  \bibinfo{person}{N. Lu}, {and} \bibinfo{person}{D.~A. Frincke}.}
  \bibinfo{year}{2010}\natexlab{}.
\newblock \showarticletitle{Smart-grid security issues}.
\newblock \bibinfo{journal}{{\em IEEE Security Privacy\/}} \bibinfo{volume}{8},
  \bibinfo{number}{1} (\bibinfo{date}{Jan} \bibinfo{year}{2010}),
  \bibinfo{pages}{81--85}.
\newblock
\showISSN{1540-7993}
\showDOI{%
\url{https://doi.org/10.1109/MSP.2010.49}}


\bibitem[\protect\citeauthoryear{Kim and Zeldovich}{Kim and Zeldovich}{2013}]%
        {interposition_2}
\bibfield{author}{\bibinfo{person}{T. Kim} {and} \bibinfo{person}{N.
  Zeldovich}.} \bibinfo{year}{2013}\natexlab{}.
\newblock \showarticletitle{Practical and Effective Sandboxing for Non-root
  Users}. In \bibinfo{booktitle}{{\em Proceedings of the 2013 USENIX Conference
  on Annual Technical Conference}} {\em (\bibinfo{series}{USENIX ATC'13})}.
  \bibinfo{publisher}{USENIX Association}, \bibinfo{address}{Berkeley, CA,
  USA}, \bibinfo{pages}{139--144}.
\newblock
\showURL{%
\url{http://dl.acm.org/citation.cfm?id=2535461.2535478}}


\bibitem[\protect\citeauthoryear{Kosek}{Kosek}{2016}]%
        {anomalydetection}
\bibfield{author}{\bibinfo{person}{A.~M. Kosek}.}
  \bibinfo{year}{2016}\natexlab{}.
\newblock \showarticletitle{Contextual anomaly detection for cyber-physical
  security in Smart Grids based on an artificial neural network model}. In
  \bibinfo{booktitle}{{\em 2016 Joint Workshop on Cyber- Physical Security and
  Resilience in Smart Grids (CPSR-SG)}}. \bibinfo{pages}{1--6}.
\newblock
\showDOI{%
\url{https://doi.org/10.1109/CPSRSG.2016.7684103}}


\bibitem[\protect\citeauthoryear{Kushner}{Kushner}{2013}]%
        {stuxnetstory}
\bibfield{author}{\bibinfo{person}{D. Kushner}.}
  \bibinfo{year}{2013}\natexlab{}.
\newblock \showarticletitle{The real story of stuxnet}.
\newblock \bibinfo{journal}{{\em IEEE Spectrum\/}} \bibinfo{volume}{50},
  \bibinfo{number}{3} (\bibinfo{date}{March} \bibinfo{year}{2013}),
  \bibinfo{pages}{48--53}.
\newblock
\showISSN{0018-9235}
\showDOI{%
\url{https://doi.org/10.1109/MSPEC.2013.6471059}}


\bibitem[\protect\citeauthoryear{{L. Babun, H. Aksu and A. S. Uluagac}}{{L.
  Babun, H. Aksu and A. S. Uluagac}}{2017}]%
        {ICCPaper}
\bibfield{author}{\bibinfo{person}{{L. Babun, H. Aksu and A. S. Uluagac}}.}
  \bibinfo{year}{2017}\natexlab{}.
\newblock \showarticletitle{{Identifying Counterfeit Smart Grid Devices: A
  Lightweight System Level Framework}}. In \bibinfo{booktitle}{{\em
  {Proceedings of the IEEE ICC Intern. Conf. on Communications}}}.
  \bibinfo{publisher}{IEEE}, \bibinfo{address}{Paris, France}, 7.
\newblock


\bibitem[\protect\citeauthoryear{Lago, Ferrante, Passerone, and Ferrari}{Lago
  et~al\mbox{.}}{2018}]%
        {modelling1}
\bibfield{author}{\bibinfo{person}{L.~D. Lago}, \bibinfo{person}{O. Ferrante},
  \bibinfo{person}{R. Passerone}, {and} \bibinfo{person}{A. Ferrari}.}
  \bibinfo{year}{2018}\natexlab{}.
\newblock \showarticletitle{Dependability Assessment of SOA-Based CPS With
  Contracts and Model-Based Fault Injection}.
\newblock \bibinfo{journal}{{\em IEEE Transactions on Industrial
  Informatics\/}} \bibinfo{volume}{14}, \bibinfo{number}{1}
  (\bibinfo{date}{Jan} \bibinfo{year}{2018}), \bibinfo{pages}{360--369}.
\newblock
\showISSN{1551-3203}
\showDOI{%
\url{https://doi.org/10.1109/TII.2017.2689337}}


\bibitem[\protect\citeauthoryear{{LeMay}, {Gross}, {Gunter}, and
  {Garg}}{{LeMay} et~al\mbox{.}}{2007}]%
        {attestationHW}
\bibfield{author}{\bibinfo{person}{M. {LeMay}}, \bibinfo{person}{G. {Gross}},
  \bibinfo{person}{C.~A. {Gunter}}, {and} \bibinfo{person}{S. {Garg}}.}
  \bibinfo{year}{2007}\natexlab{}.
\newblock \showarticletitle{Unified Architecture for Large-Scale Attested
  Metering}. In \bibinfo{booktitle}{{\em 2007 40th Annual Hawaii International
  Conference on System Sciences (HICSS'07)}}. \bibinfo{pages}{115--115}.
\newblock
\showISSN{1530-1605}
\showDOI{%
\url{https://doi.org/10.1109/HICSS.2007.586}}


\bibitem[\protect\citeauthoryear{Lopez, Babun, Aksu, and Uluagac}{Lopez
  et~al\mbox{.}}{2017}]%
        {hookingJournal}
\bibfield{author}{\bibinfo{person}{Juan Lopez}, \bibinfo{person}{Leonardo
  Babun}, \bibinfo{person}{Hidayet Aksu}, {and} \bibinfo{person}{A.~Selcuk
  Uluagac}.} \bibinfo{year}{2017}\natexlab{}.
\newblock \showarticletitle{A Survey on Function and System Call Hooking
  Approaches}.
\newblock \bibinfo{journal}{{\em Journal of Hardware and Systems Security\/}}
  \bibinfo{volume}{1}, \bibinfo{number}{2} (\bibinfo{date}{01 Jun}
  \bibinfo{year}{2017}), \bibinfo{pages}{114--136}.
\newblock
\showISSN{2509-3436}
\showDOI{%
\url{https://doi.org/10.1007/s41635-017-0013-2}}


\bibitem[\protect\citeauthoryear{Loscocco, Wilson, Pendergrass, and
  McDonell}{Loscocco et~al\mbox{.}}{2007}]%
        {lkim}
\bibfield{author}{\bibinfo{person}{Peter~A. Loscocco},
  \bibinfo{person}{Perry~W. Wilson}, \bibinfo{person}{J.~Aaron Pendergrass},
  {and} \bibinfo{person}{C.~Durward McDonell}.}
  \bibinfo{year}{2007}\natexlab{}.
\newblock \showarticletitle{Linux Kernel Integrity Measurement Using Contextual
  Inspection}. In \bibinfo{booktitle}{{\em Proceedings of the 2007 ACM Workshop
  on Scalable Trusted Computing}} {\em (\bibinfo{series}{STC '07})}.
  \bibinfo{publisher}{ACM}, \bibinfo{address}{New York, NY, USA},
  \bibinfo{pages}{21--29}.
\newblock
\showISBNx{978-1-59593-888-6}
\showDOI{%
\url{https://doi.org/10.1145/1314354.1314362}}


\bibitem[\protect\citeauthoryear{{M. Q. Saeed, Z. Bilal and C. D. Walter}}{{M.
  Q. Saeed, Z. Bilal and C. D. Walter}}{2013}]%
        {supplychainproblem}
\bibfield{author}{\bibinfo{person}{{M. Q. Saeed, Z. Bilal and C. D. Walter}}.}
  \bibinfo{year}{2013}\natexlab{}.
\newblock \showarticletitle{{An NFC based onsumer-level counterfeit detection
  framework}}. In \bibinfo{booktitle}{{\em 2013 Eleventh Annual Int. Conf. on
  Privacy, Security and Trust (PST)}}. \bibinfo{publisher}{IEEE},
  \bibinfo{address}{Tarragona}, \bibinfo{pages}{135--142}.
\newblock


\bibitem[\protect\citeauthoryear{{M, Sillgith}}{{M, Sillgith}}{2016}]%
        {libiec61850}
\bibfield{author}{\bibinfo{person}{{M, Sillgith}}.}
  \bibinfo{year}{2016}\natexlab{}.
\newblock \bibinfo{title}{{Open source library for IEC 61850: Release 0.9}}.
\newblock   (\bibinfo{date}{Feb} \bibinfo{year}{2016}).
\newblock
\showURL{%
\url{http://libiec61850.com/libiec61850/}}


\bibitem[\protect\citeauthoryear{Milea, Khoo, Lo, and Pop}{Milea
  et~al\mbox{.}}{2012}]%
        {systemcalls3}
\bibfield{author}{\bibinfo{person}{Narcisa~Andreea Milea},
  \bibinfo{person}{Siau~Cheng Khoo}, \bibinfo{person}{David Lo}, {and}
  \bibinfo{person}{Cristian Pop}.} \bibinfo{year}{2012}\natexlab{}.
\newblock \showarticletitle{NORT: Runtime Anomaly-based Monitoring of Malicious
  Behavior for Windows}. In \bibinfo{booktitle}{{\em Proceedings of the Second
  International Conference on Runtime Verification}} {\em
  (\bibinfo{series}{RV'11})}. \bibinfo{publisher}{Springer-Verlag},
  \bibinfo{address}{Berlin, Heidelberg}, \bibinfo{pages}{115--130}.
\newblock
\showISBNx{978-3-642-29859-2}
\showDOI{%
\url{https://doi.org/10.1007/978-3-642-29860-8_10}}


\bibitem[\protect\citeauthoryear{{Murguia} and {Ruths}}{{Murguia} and
  {Ruths}}{2016}]%
        {cusum}
\bibfield{author}{\bibinfo{person}{C. {Murguia}} {and} \bibinfo{person}{J.
  {Ruths}}.} \bibinfo{year}{2016}\natexlab{}.
\newblock \showarticletitle{CUSUM and chi-squared attack detection of
  compromised sensors}. In \bibinfo{booktitle}{{\em 2016 IEEE Conference on
  Control Applications (CCA)}}. \bibinfo{pages}{474--480}.
\newblock
\showDOI{%
\url{https://doi.org/10.1109/CCA.2016.7587875}}


\bibitem[\protect\citeauthoryear{{N. Komninos, E. Philippou and A.
  Pitsillides}}{{N. Komninos, E. Philippou and A. Pitsillides}}{2014}]%
        {SEC_grid_home}
\bibfield{author}{\bibinfo{person}{{N. Komninos, E. Philippou and A.
  Pitsillides}}.} \bibinfo{year}{2014}\natexlab{}.
\newblock \showarticletitle{{Survey in smart grid and smart home security:
  issues, challenges and countermeasures}}.
\newblock \bibinfo{journal}{{\em IEEE Communications Surveys and Tutorials\/}}
  \bibinfo{volume}{16} (\bibinfo{year}{2014}), \bibinfo{pages}{1933--1954}.
\newblock
Issue 4.
\showDOI{%
\url{https://doi.org/10.1109/COMST.2014.2320093}}


\bibitem[\protect\citeauthoryear{{National Cybersecurity \& Communications
  Integration Center (NCCIC), Department of Homeland Security}}{{National
  Cybersecurity \& Communications Integration Center (NCCIC), Department of
  Homeland Security}}{2018}]%
        {russian}
\bibfield{author}{\bibinfo{person}{{National Cybersecurity \& Communications
  Integration Center (NCCIC), Department of Homeland Security}}.}
  \bibinfo{year}{2018}\natexlab{}.
\newblock \bibinfo{title}{Russian Activity Against Critical Infrastructure}.
\newblock   (\bibinfo{year}{2018}).
\newblock
\showURL{%
\url{https://www.us-cert.gov/sites/default/files/c3vp/Russian_Activity_Webinar_Slides.pdf}}


\bibitem[\protect\citeauthoryear{{NIST Special Publication 1108r3}}{{NIST
  Special Publication 1108r3}}{2014}]%
        {NIST_framework3}
\bibfield{author}{\bibinfo{person}{{NIST Special Publication 1108r3}}.}
  \bibinfo{year}{2014}\natexlab{}.
\newblock \bibinfo{title}{{NIST framework and roadmap for smart grid
  interoperability standards, release 3.0}}.
\newblock   (\bibinfo{date}{Sep} \bibinfo{year}{2014}).
\newblock
\showURL{%
\url{http://nvlpubs.nist.gov/nistpubs/SpecialPublications/NIST.SP.1108r3.pdf}}


\bibitem[\protect\citeauthoryear{Nourian and Madnick}{Nourian and
  Madnick}{2015}]%
        {stuxnet2}
\bibfield{author}{\bibinfo{person}{A. Nourian} {and} \bibinfo{person}{S.
  Madnick}.} \bibinfo{year}{2015}\natexlab{}.
\newblock \showarticletitle{A Systems Theoretic Approach to the Security
  Threats in Cyber Physical Systems Applied to Stuxnet}.
\newblock \bibinfo{journal}{{\em IEEE Transactions on Dependable and Secure
  Computing\/}} \bibinfo{volume}{PP}, \bibinfo{number}{99}
  (\bibinfo{year}{2015}), \bibinfo{pages}{1--1}.
\newblock
\showISSN{1545-5971}
\showDOI{%
\url{https://doi.org/10.1109/TDSC.2015.2509994}}


\bibitem[\protect\citeauthoryear{Ozay, Esnaola, Vural, Kulkarni, and Poor}{Ozay
  et~al\mbox{.}}{2016}]%
        {attackdetection5}
\bibfield{author}{\bibinfo{person}{M. Ozay}, \bibinfo{person}{I. Esnaola},
  \bibinfo{person}{F.~T.~Yarman Vural}, \bibinfo{person}{S.~R. Kulkarni}, {and}
  \bibinfo{person}{H.~V. Poor}.} \bibinfo{year}{2016}\natexlab{}.
\newblock \showarticletitle{Machine Learning Methods for Attack Detection in
  the Smart Grid}.
\newblock \bibinfo{journal}{{\em IEEE Transactions on Neural Networks and
  Learning Systems\/}} \bibinfo{volume}{27}, \bibinfo{number}{8}
  (\bibinfo{date}{Aug} \bibinfo{year}{2016}), \bibinfo{pages}{1773--1786}.
\newblock
\showISSN{2162-237X}
\showDOI{%
\url{https://doi.org/10.1109/TNNLS.2015.2404803}}


\bibitem[\protect\citeauthoryear{{Pendergrass}, {Helble}, {Clemens}, and
  {Loscocco}}{{Pendergrass} et~al\mbox{.}}{2018}]%
        {maat}
\bibfield{author}{\bibinfo{person}{J.~A. {Pendergrass}}, \bibinfo{person}{S.
  {Helble}}, \bibinfo{person}{J. {Clemens}}, {and} \bibinfo{person}{P.
  {Loscocco}}.} \bibinfo{year}{2018}\natexlab{}.
\newblock \showarticletitle{A Platform Service for Remote Integrity Measurement
  and Attestation}. In \bibinfo{booktitle}{{\em MILCOM 2018 - 2018 IEEE
  Military Communications Conference (MILCOM)}}. \bibinfo{pages}{1--6}.
\newblock
\showISSN{2155-7586}
\showDOI{%
\url{https://doi.org/10.1109/MILCOM.2018.8599735}}


\bibitem[\protect\citeauthoryear{Pendleton and Xu}{Pendleton and Xu}{2017}]%
        {syscalls1}
\bibfield{author}{\bibinfo{person}{M. Pendleton} {and} \bibinfo{person}{S.
  Xu}.} \bibinfo{year}{2017}\natexlab{}.
\newblock \showarticletitle{A dataset generator for next generation system call
  host intrusion detection systems}. In \bibinfo{booktitle}{{\em MILCOM 2017 -
  2017 IEEE Military Communications Conference (MILCOM)}}.
  \bibinfo{pages}{231--236}.
\newblock
\showDOI{%
\url{https://doi.org/10.1109/MILCOM.2017.8170835}}


\bibitem[\protect\citeauthoryear{Rawat and Bajracharya}{Rawat and
  Bajracharya}{2015}]%
        {SEC_smart_grid_systems}
\bibfield{author}{\bibinfo{person}{D.~B. Rawat} {and} \bibinfo{person}{Ch.
  Bajracharya}.} \bibinfo{year}{2015}\natexlab{}.
\newblock \showarticletitle{{Cyber security for smart grid systems: status,
  challenges and perspectives}}. In \bibinfo{booktitle}{{\em Proceedings of the
  IEEE Southeast Conf}}. \bibinfo{publisher}{IEEE}, \bibinfo{address}{Fort
  Lauderdale, FL, USA}, \bibinfo{pages}{1--6}.
\newblock
\showDOI{%
\url{https://doi.org/10.1109/SECON.2015.7132891}}


\bibitem[\protect\citeauthoryear{{Reuters}}{{Reuters}}{2016}]%
        {reuters}
\bibfield{author}{\bibinfo{person}{{Reuters}}.}
  \bibinfo{year}{2016}\natexlab{}.
\newblock \bibinfo{title}{U.S. firm blames Russian 'Sandworm' hackers for
  Ukraine outage}.
\newblock   (\bibinfo{year}{2016}).
\newblock
\showURL{%
\url{https://www.reuters.com/article/us-ukraine-cybersecurity-sandworm/u-s-firm-blames-russian-sandworm-hackers-for-ukraine-outage-idUSKBN0UM00N20160108}}


\bibitem[\protect\citeauthoryear{Ross}{Ross}{2001}]%
        {RossBook}
\bibfield{author}{\bibinfo{person}{Sheldon~M. Ross}.}
  \bibinfo{year}{2001}\natexlab{}.
\newblock \bibinfo{booktitle}{{\em Probability Models for Computer Science\/}
  (\bibinfo{edition}{1st} ed.)}.
\newblock \bibinfo{publisher}{Academic Press, Inc.}, \bibinfo{address}{Orlando,
  FL, USA}.
\newblock
\showISBNx{0125980515}


\bibitem[\protect\citeauthoryear{{S. Fries, H. J. Hof and M. G. Seewald}}{{S.
  Fries, H. J. Hof and M. G. Seewald}}{2010}]%
        {SEC_enhancing62351}
\bibfield{author}{\bibinfo{person}{{S. Fries, H. J. Hof and M. G. Seewald}}.}
  \bibinfo{year}{2010}\natexlab{}.
\newblock \showarticletitle{{Security of the smart grid - enhancing {IEC} 62351
  to improve security in energy automation control}}.
\newblock \bibinfo{journal}{{\em Int. Journal on Advances in Security\/}}
  \bibinfo{volume}{3} (\bibinfo{year}{2010}).
\newblock
Issue 3, 4.
\showDOI{%
\url{https://doi.org/10.1.1.474.6536}}


\bibitem[\protect\citeauthoryear{Sanjab and Saad}{Sanjab and Saad}{2016}]%
        {datainjection1}
\bibfield{author}{\bibinfo{person}{A. Sanjab} {and} \bibinfo{person}{W. Saad}.}
  \bibinfo{year}{2016}\natexlab{}.
\newblock \showarticletitle{Data Injection Attacks on Smart Grids With Multiple
  Adversaries: A Game-Theoretic Perspective}.
\newblock \bibinfo{journal}{{\em IEEE Transactions on Smart Grid\/}}
  \bibinfo{volume}{7}, \bibinfo{number}{4} (\bibinfo{date}{July}
  \bibinfo{year}{2016}), \bibinfo{pages}{2038--2049}.
\newblock
\showISSN{1949-3053}
\showDOI{%
\url{https://doi.org/10.1109/TSG.2016.2550218}}


\bibitem[\protect\citeauthoryear{Sanjab, Saad, G{\"{u}}ven{\c{c}}, Sarwat, and
  Biswas}{Sanjab et~al\mbox{.}}{2016}]%
        {compromised1}
\bibfield{author}{\bibinfo{person}{Anibal Sanjab}, \bibinfo{person}{Walid
  Saad}, \bibinfo{person}{Ismail G{\"{u}}ven{\c{c}}}, \bibinfo{person}{Arif~I.
  Sarwat}, {and} \bibinfo{person}{Saroj Biswas}.}
  \bibinfo{year}{2016}\natexlab{}.
\newblock \showarticletitle{Smart Grid Security: Threats, Challenges, and
  Solutions}.
\newblock \bibinfo{journal}{{\em CoRR\/}}  \bibinfo{volume}{abs/1606.06992}
  (\bibinfo{year}{2016}).
\newblock
\showURL{%
\url{http://arxiv.org/abs/1606.06992}}


\bibitem[\protect\citeauthoryear{Sedjelmaci and Senouci}{Sedjelmaci and
  Senouci}{2016}]%
        {compromised2}
\bibfield{author}{\bibinfo{person}{H. Sedjelmaci} {and} \bibinfo{person}{S.~M.
  Senouci}.} \bibinfo{year}{2016}\natexlab{}.
\newblock \showarticletitle{Smart grid Security: A new approach to detect
  intruders in a smart grid Neighborhood Area Network}. In
  \bibinfo{booktitle}{{\em 2016 International Conference on Wireless Networks
  and Mobile Communications (WINCOM)}}. \bibinfo{pages}{6--11}.
\newblock
\showDOI{%
\url{https://doi.org/10.1109/WINCOM.2016.7777182}}


\bibitem[\protect\citeauthoryear{Sekar, Bendre, Dhurjati, and Bollineni}{Sekar
  et~al\mbox{.}}{2001}]%
        {systemcalls1}
\bibfield{author}{\bibinfo{person}{R. Sekar}, \bibinfo{person}{M. Bendre},
  \bibinfo{person}{D. Dhurjati}, {and} \bibinfo{person}{P. Bollineni}.}
  \bibinfo{year}{2001}\natexlab{}.
\newblock \showarticletitle{A Fast Automaton-Based Method for Detecting
  Anomalous Program Behaviors}. In \bibinfo{booktitle}{{\em Proceedings of the
  2001 IEEE Symposium on Security and Privacy}} {\em (\bibinfo{series}{SP
  '01})}. \bibinfo{publisher}{IEEE Computer Society},
  \bibinfo{address}{Washington, DC, USA}, \bibinfo{pages}{144--}.
\newblock
\showURL{%
\url{http://dl.acm.org/citation.cfm?id=882495.884433}}


\bibitem[\protect\citeauthoryear{Sikdar and Chow}{Sikdar and Chow}{2011}]%
        {Phasor}
\bibfield{author}{\bibinfo{person}{B. Sikdar} {and} \bibinfo{person}{J.~H.
  Chow}.} \bibinfo{year}{2011}\natexlab{}.
\newblock \showarticletitle{{Defending synchrophasor data networks against
  traffic analysis attacks.}}
\newblock \bibinfo{journal}{{\em IEEE Transactions on Smart Grid\/}}
  \bibinfo{volume}{2} (\bibinfo{year}{2011}), \bibinfo{pages}{819--826}.
\newblock
Issue 4.


\bibitem[\protect\citeauthoryear{Sun, Guan, Liu, and Liu}{Sun
  et~al\mbox{.}}{2013}]%
        {attackdetection1}
\bibfield{author}{\bibinfo{person}{Y. Sun}, \bibinfo{person}{X. Guan},
  \bibinfo{person}{T. Liu}, {and} \bibinfo{person}{Y. Liu}.}
  \bibinfo{year}{2013}\natexlab{}.
\newblock \showarticletitle{A cyber-physical monitoring system for attack
  detection in smart grid}. In \bibinfo{booktitle}{{\em 2013 IEEE Conference on
  Computer Communications Workshops (INFOCOM WKSHPS)}}.
  \bibinfo{pages}{33--34}.
\newblock
\showDOI{%
\url{https://doi.org/10.1109/INFCOMW.2013.6970712}}


\bibitem[\protect\citeauthoryear{{Symantec}}{{Symantec}}{2018}]%
        {indirectattack}
\bibfield{author}{\bibinfo{person}{{Symantec}}.}
  \bibinfo{year}{2018}\natexlab{}.
\newblock \bibinfo{title}{Sandworm Windows zero-day vulnerability being
  actively exploited in targeted attacks}.
\newblock   (\bibinfo{year}{2018}).
\newblock
\showURL{%
\url{https://www.symantec.com/connect/blogs/sandworm-windows-zero-day-vulnerability-being-actively-exploited-targeted-attacks}}


\bibitem[\protect\citeauthoryear{Tazi, Abdi, and Abbou}{Tazi
  et~al\mbox{.}}{2015}]%
        {securityissues2}
\bibfield{author}{\bibinfo{person}{K. Tazi}, \bibinfo{person}{F. Abdi}, {and}
  \bibinfo{person}{M.~F. Abbou}.} \bibinfo{year}{2015}\natexlab{}.
\newblock \showarticletitle{Review on cyber-physical security of the smart
  grid: Attacks and defense mechanisms}. In \bibinfo{booktitle}{{\em 2015 3rd
  International Renewable and Sustainable Energy Conference (IRSEC)}}.
  \bibinfo{pages}{1--6}.
\newblock
\showDOI{%
\url{https://doi.org/10.1109/IRSEC.2015.7455127}}


\bibitem[\protect\citeauthoryear{{The smart grid interoperability panel - cyber
  security working group}}{{The smart grid interoperability panel - cyber
  security working group}}{2010}]%
        {NIST_security_panel}
\bibfield{author}{\bibinfo{person}{{The smart grid interoperability panel -
  cyber security working group}}.} \bibinfo{year}{2010}\natexlab{}.
\newblock \bibinfo{title}{{Introduction to NISTIR 7628: guidelines for smart
  grid cyber security}}.
\newblock   (\bibinfo{date}{Sept} \bibinfo{year}{2010}).
\newblock
\showURL{%
\url{http://www.nist.gov/smartgrid/upload/nistir-7628_total.pdf}}


\bibitem[\protect\citeauthoryear{{Valente}, {Barreto}, and
  {C\'{a}rdenas}}{{Valente} et~al\mbox{.}}{2014}]%
        {attestation}
\bibfield{author}{\bibinfo{person}{J. {Valente}}, \bibinfo{person}{C.
  {Barreto}}, {and} \bibinfo{person}{A.~A. {C\'{a}rdenas}}.}
  \bibinfo{year}{2014}\natexlab{}.
\newblock \showarticletitle{Cyber-Physical Systems Attestation}. In
  \bibinfo{booktitle}{{\em 2014 IEEE International Conference on Distributed
  Computing in Sensor Systems}}. \bibinfo{pages}{354--357}.
\newblock
\showISSN{2325-2936}
\showDOI{%
\url{https://doi.org/10.1109/DCOSS.2014.61}}


\bibitem[\protect\citeauthoryear{Wang and Lu}{Wang and Lu}{2013}]%
        {SEC_grid_challenges}
\bibfield{author}{\bibinfo{person}{W. Wang} {and} \bibinfo{person}{Z. Lu}.}
  \bibinfo{year}{2013}\natexlab{}.
\newblock \showarticletitle{Survey Cyber Security in the Smart Grid: Survey and
  Challenges}.
\newblock \bibinfo{journal}{{\em Comput. Netw.\/}} \bibinfo{volume}{57},
  \bibinfo{number}{5} (\bibinfo{date}{April} \bibinfo{year}{2013}),
  \bibinfo{pages}{1344--1371}.
\newblock
\showISSN{1389-1286}
\showDOI{%
\url{https://doi.org/10.1016/j.comnet.2012.12.017}}


\bibitem[\protect\citeauthoryear{{X. Li, I. Lille, X. Liang, R. Lu, X. Shen, X.
  Lin and H. Zhu}}{{X. Li, I. Lille, X. Liang, R. Lu, X. Shen, X. Lin and H.
  Zhu}}{2012}]%
        {SEC_cyber_attacks}
\bibfield{author}{\bibinfo{person}{{X. Li, I. Lille, X. Liang, R. Lu, X. Shen,
  X. Lin and H. Zhu}}.} \bibinfo{year}{2012}\natexlab{}.
\newblock \showarticletitle{{Securing smart grid: cyber attacks,
  countermeasures and challenges}}.
\newblock \bibinfo{journal}{{\em IEEE Comm. magazine\/}}  \bibinfo{volume}{50}
  (\bibinfo{year}{2012}), \bibinfo{pages}{38--45}.
\newblock
Issue 8.
\showDOI{%
\url{https://doi.org/10.1109/MCOM.2012.6257525}}


\bibitem[\protect\citeauthoryear{{Y. Obeng, C. Nolan and D. Brown}}{{Y. Obeng,
  C. Nolan and D. Brown}}{2016}]%
        {supplychainproblem2}
\bibfield{author}{\bibinfo{person}{{Y. Obeng, C. Nolan and D. Brown}}.}
  \bibinfo{year}{2016}\natexlab{}.
\newblock \showarticletitle{{Hardware security through chain assurance}}. In
  \bibinfo{booktitle}{{\em Design, Automation and Test in Europe Conf. and
  Exhibition (DATE)}}. \bibinfo{publisher}{IEEE}, \bibinfo{address}{Dresden},
  \bibinfo{pages}{1535 -- 1537}.
\newblock


\bibitem[\protect\citeauthoryear{{Y. Yan, Y. Qian, H. Sharif and D.
  Tipper}}{{Y. Yan, Y. Qian, H. Sharif and D. Tipper}}{2012}]%
        {SEC_grid_comm}
\bibfield{author}{\bibinfo{person}{{Y. Yan, Y. Qian, H. Sharif and D.
  Tipper}}.} \bibinfo{year}{2012}\natexlab{}.
\newblock \showarticletitle{{A survey on cyber security for smart grid
  communications}}.
\newblock \bibinfo{journal}{{\em IEEE Communications Surveys and Tutorials\/}}
  \bibinfo{volume}{14} (\bibinfo{year}{2012}), \bibinfo{pages}{998--1010}.
\newblock
Issue 4.
\showDOI{%
\url{https://doi.org/10.1109/SURV.2012.010912.00035}}


\bibitem[\protect\citeauthoryear{Yang, Crossley, Wen, Chatfield, and
  Wright}{Yang et~al\mbox{.}}{2014}]%
        {tsg}
\bibfield{author}{\bibinfo{person}{L. Yang}, \bibinfo{person}{P.~A. Crossley},
  \bibinfo{person}{A. Wen}, \bibinfo{person}{R. Chatfield}, {and}
  \bibinfo{person}{J. Wright}.} \bibinfo{year}{2014}\natexlab{}.
\newblock \showarticletitle{Design and Performance Testing of a Multivendor
  IEC61850 \#x2013;9-2 Process Bus Based Protection Scheme}.
\newblock \bibinfo{journal}{{\em IEEE Transactions on Smart Grid\/}}
  \bibinfo{volume}{5}, \bibinfo{number}{3} (\bibinfo{date}{May}
  \bibinfo{year}{2014}), \bibinfo{pages}{1159--1164}.
\newblock
\showISSN{1949-3053}
\showDOI{%
\url{https://doi.org/10.1109/TSG.2013.2277940}}


\bibitem[\protect\citeauthoryear{Yang, Min, An, Yu, and Yang}{Yang
  et~al\mbox{.}}{2016}]%
        {attackdetection3}
\bibfield{author}{\bibinfo{person}{Q. Yang}, \bibinfo{person}{Rui Min},
  \bibinfo{person}{D. An}, \bibinfo{person}{W. Yu}, {and} \bibinfo{person}{X.
  Yang}.} \bibinfo{year}{2016}\natexlab{}.
\newblock \showarticletitle{{Towards optimal PMU placement against data
  integrity attacks in smart grid}}. In \bibinfo{booktitle}{{\em 2016 Annual
  Conference on Information Science and Systems (CISS)}}.
  \bibinfo{pages}{54--58}.
\newblock
\showDOI{%
\url{https://doi.org/10.1109/CISS.2016.7460476}}


\bibitem[\protect\citeauthoryear{Zhou and Makris}{Zhou and Makris}{2017}]%
        {syscall2}
\bibfield{author}{\bibinfo{person}{L. Zhou} {and} \bibinfo{person}{Y. Makris}.}
  \bibinfo{year}{2017}\natexlab{}.
\newblock \showarticletitle{Hardware-based on-line intrusion detection via
  system call routine fingerprinting}. In \bibinfo{booktitle}{{\em Design,
  Automation Test in Europe Conference Exhibition (DATE), 2017}}.
  \bibinfo{pages}{1546--1551}.
\newblock
\showDOI{%
\url{https://doi.org/10.23919/DATE.2017.7927236}}


\bibitem[\protect\citeauthoryear{Zhou and Miao}{Zhou and Miao}{2016}]%
        {statestimationattack1}
\bibfield{author}{\bibinfo{person}{Y. Zhou} {and} \bibinfo{person}{Z. Miao}.}
  \bibinfo{year}{2016}\natexlab{}.
\newblock \showarticletitle{Cyber attacks, detection and protection in smart
  grid state estimation}. In \bibinfo{booktitle}{{\em 2016 North American Power
  Symposium (NAPS)}}. \bibinfo{pages}{1--6}.
\newblock
\showDOI{%
\url{https://doi.org/10.1109/NAPS.2016.7747874}}
\vspace{0.5in}

\end{thebibliography}


%
\begin{wrapfigure}{l}{.95in}
 \centering{
    \includegraphics[width=.2\textwidth]{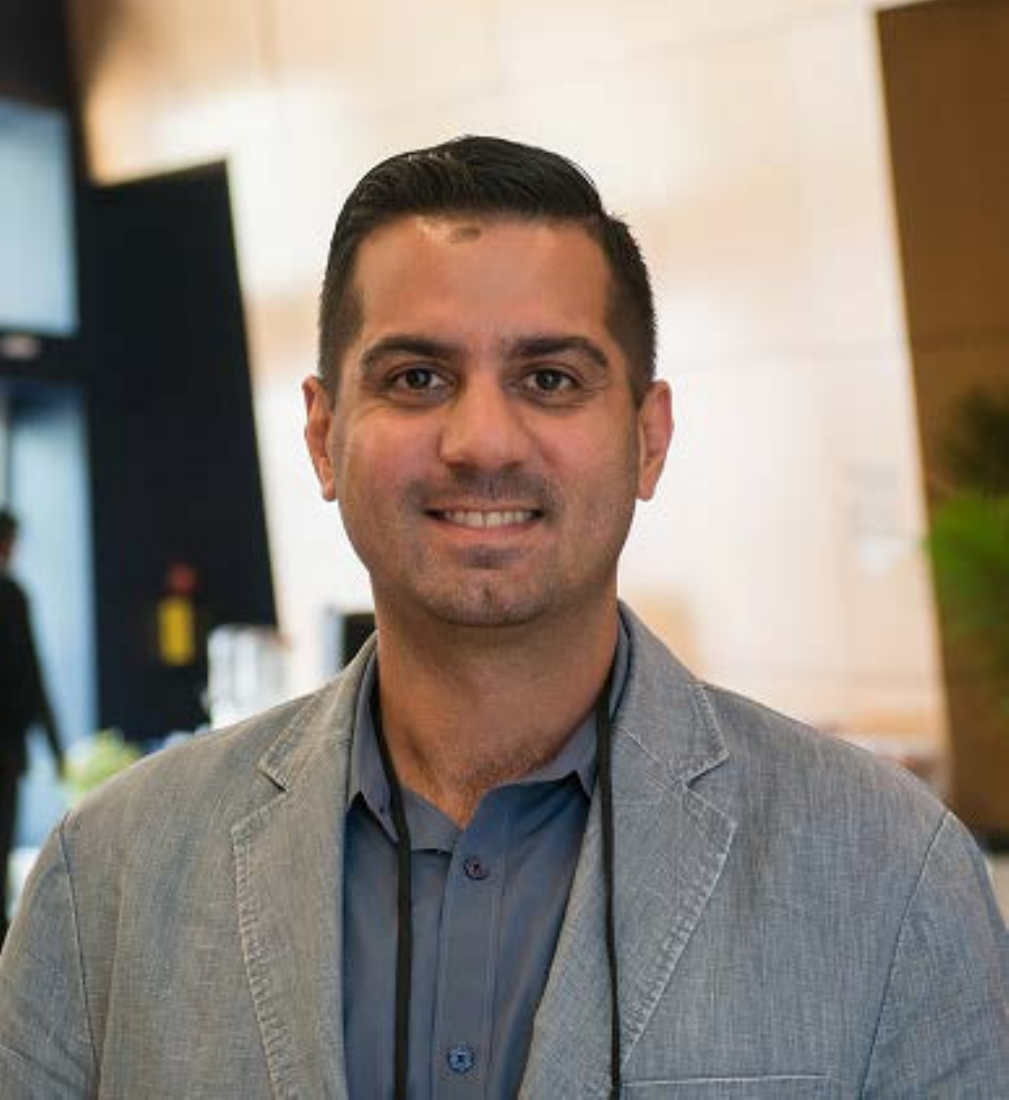}}
\end{wrapfigure} 
\textbf{Leonardo Babun} is a Ph.D. candidate and CyberCorps Scholarship for Service Fellow in the Department of Electrical \& Computer Engineering at Florida International University. He is also a member of the Cyber-Physical Systems Security Lab (CSL). Leonardo previously completed his M.S. in Electrical Engineering from the Department of Electrical \& Computer Engineering at Florida International University in 2015. His research interests are focused on the security and privacy of Cyber-Physical Systems (CPS) and the Internet of Things (IoT), digital forensics of smart settings, wireless networks, big data analytics, and distributed computing. You can contact him at lbabu002@fiu.edu. 
\vfill

\begin{wrapfigure}{l}{.95in}
    \includegraphics[width=.2\columnwidth]{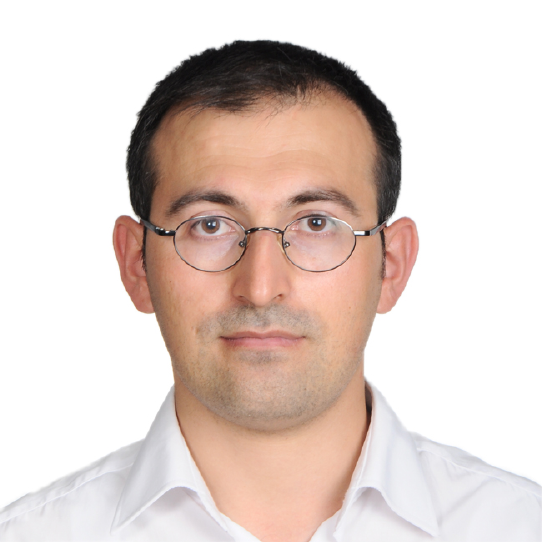}
  \end{wrapfigure} 
\textbf{Hidayet Aksu} received his Ph.D., M.S. and B.S. degrees from Bilkent University, all in Department of Computer Engineering, in 2014, 2008 and 2005, respectively. He is currently a Postdoctoral Associate in the Department of Electrical \& Computer Engineering at Florida International University (FIU). Before that, he worked as an Adjunct Faculty in the Computer Engineering Department of Bilkent University. He conducted research as visiting scholar at IBM T.J. Watson Research Center, USA in 2012-2013. He also worked for Scientific and Technological Research Council of Turkey (TUBITAK).
His research interests include security for cyber-physical systems, internet of things, security for critical infrastructure networks, IoT security, security analytics, social networks, big data analytics, distributed computing, wireless networks, wireless ad hoc and sensor networks, localization, and P2P networks.

\begin{wrapfigure}{l}{.95in}
    \includegraphics[width=.2\columnwidth]{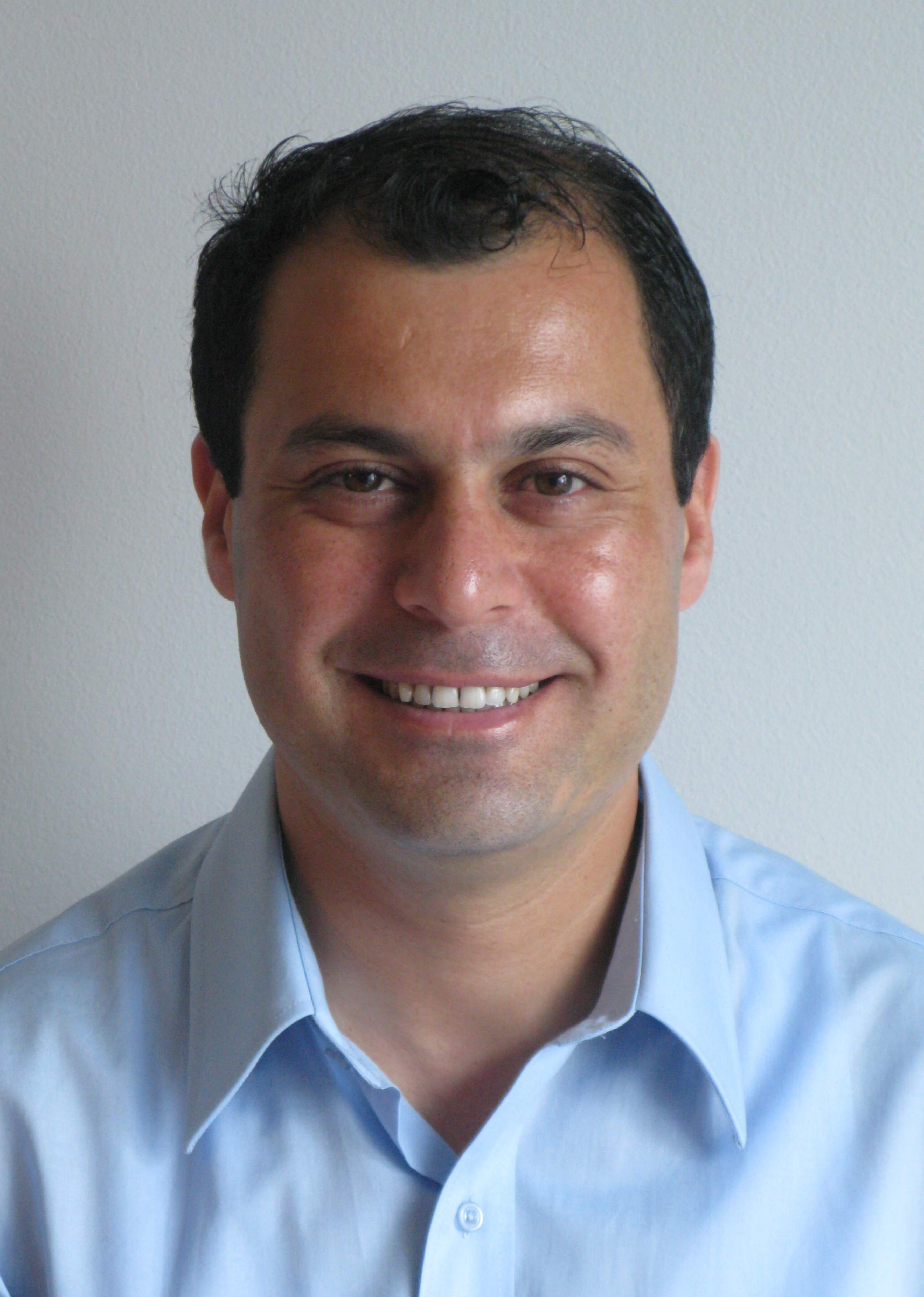}
  \end{wrapfigure} 
  
\vfill 
\textbf{Selcuk Uluagac} leads the Cyber-Physical Systems Security Lab at Florida International University, focusing on security and privacy of Internet of Things and Cyber-Physical Systems. He has a Ph.D. and M.S. from Georgia Institute of Technology, and M.S. from Carnegie Mellon University. In 2015, he received the US National Science Foundation CAREER award and US Air Force Office of Sponsored Research’s Summer Faculty Fellowship, and in 2016, Summer Faculty Fellowship from University of Padova, Italy. Currently, he serves on the editorial boards of Elsevier Journal of Network and Computer Applications (JNCA), Elsevier Computer Networks,  and the IEEE Communications and Surveys and Tutorials. 

\end{document}